\documentstyle[eqsecnum,aps,pra,graphicx,floats]{revtex}

\newcommand{\bq}{\begin{equation}}
\newcommand{\eq}{\end{equation}}
\newcommand{\th}{\theta}
\newcommand{\bqa}{\begin{eqnarray}}
\newcommand{\eqa}{\end{eqnarray}}
\newcommand{\nn}{\nonumber}
\renewcommand{\l}{\lambda}
\newcommand{\ket}[1]{|#1\rangle}
\newcommand{\bra}[1]{\langle#1|}
\newcommand{\R}{{\cal R}}
\newcommand{\RI}{\R^{-1}}
\newcommand{\U}{U}
\newcommand{\Tr}{\text{Tr}}
\renewcommand{\Re}{\,\text{Re}\!}
\newcommand{\bc}{\begin{center}}
\newcommand{\ec}{\end{center}}

\begin{document}
\widetext

%
%
%

\begin{flushright} { ITFA-01-14} \vspace*{7mm}\end{flushright}
\begin{center}
{\LARGE \bf Inequivalent classes of interference experiments with non-abelian anyons} \\[2ex]
{\Large B.J. Overbosch\footnote{e-mail: {\tt overbosc@science.uva.nl}} 
and F.A. Bais\footnote{e-mail: {\tt bais@science.uva.nl}}} \\[1ex]
{\em Institute for Theoretical Physics\\ University of Amsterdam\\  
     Valckenierstraat 65, 1018 XE Amsterdam\\ The Netherlands}\\
May 2001 
\end{center}
\renewcommand{\thefootnote}{\arabic{footnote}}
\setcounter{footnote}{0}

\begin{abstract}
We present a theoretical analysis of inequivalent classes of
interference experiments with non-abelian anyons using an idealized
Mach-Zender type interferometer. Because of the non-abelian nature of
the braid group action one has to distinguish the different
possibilities in which the experiment can be repeated, which lead to
different interference patterns. We show that each setup will, after
repeated measurement, lead to a situation where the two-particle (or
multi-particle) state
gets locked into an eigenstate of some well defined operator. Also the
probability to end up in such an eigenstate is calculated. Some
representative examples are worked out in detail.
\end{abstract}


\section{Introduction}\label{sec:intro}
Physics in lower dimensions is not always simpler. A nontrivial
instance in two dimensions is the possibility of {\em anyons},
i.e. excitations which exhibit statistics properties which transcend
the conventional possibilities of bosons and fermions
\cite{wilczek}. It is striking that such exotic theoretical
possibilities appear to be realized in nature. With the experimental
advances in condensed matter physics related to the fractional Quantum
Hall effect, quasi-particles have been discovered that appear to
exhibit anyonic behavior.

There is also a more mundane interest in the two dimensional physics of
anyons in the emerging field of quantum computation. In quantum
computation, as the number of qubits grows, the problem of beating
decoherence becomes a major threat: the larger the quantum system to
be manipulated gets, the more it interacts with the (noisy)
environment, causing the quantum state to collapse. In order to
compensate the inaccuracies caused by decoherence, the idea of
fault-tolerance and error-correction has become an important subject,
and quantum codes were devised where single qubit-information is
encoded into multiple qubits \cite{decoherence}. If errors in the
physical qubits can be 
corrected in time, the stored quantum information remains intact, or
at least the decoherence time of the encoded qubit is increased. The
progress achieved in the theory of quantum error codes however is
mainly mathematical in nature. Current experimental realizations of this
idea of qubit-encoding do not yet share the optimism of the theorists
and are still far from 
being decoherence free.

It may therefore be profitable to go back to some basic physics and
search for intrinsically decoherence free quantum states. Promising
candidates are ``global quantum states'' carrying quantum information
that is not `stored' locally, in single spins for instance, but in a
robust collective property of a multi-particle system -- a topological
feature for example. Such a global characteristic of the state cannot
be destroyed by local interactions, especially not by those with the
noisy environment. As anyons exhibit such a global, topological
robustness as a basic property, it seems anyons are especially suited
for this kind of the
quantum computer game, which goes by names as topological quantum
computation \cite{preskill} and geometrical quantum computation
\cite{ekert1}.

Anyons for quantum computation can be divided in two groups: the
(conventional) abelian anyons which are manipulated by mere geometric
(abelian) \emph{phases}, as in \cite{ekert1,lloyd}, and, most
promising, the so-called \emph{non-abelian anyons} (which are strictly
speaking a generalization of the abelian anyons). 
Non-abelian anyons may carry non-trivial internal degrees of freedom
which can have \emph{non-abelian topological interactions} which allow
(in principle) for a rather clean manipulation of entangled
states. {\em These non-abelian anyons may serve as intrinsically
fault-tolerant decoherence free
qubits}\cite{preskill,kitaev,freedman1,freedman2}. One might think of
non-abelian anyons in particular media with for example broken
symmetries which support quasi-particle excitations with anyonic
properties as in model systems like the so-called discrete gauge
theories \cite{bais0,bais1,bais2}.
 
A sometimes underestimated issue but necessary requirement for quantum
computation is the ability to perform a measurement of the quantum
state. In \cite{kitaev,freedman1,freedman2}, the so-called
fusion\footnote{In the discrete gauge theories \cite{bais1,bais2} the
quasi-particles are allowed to fuse and their fusion-product is a gauge
invariant, and therefore likely to be a measurable
observable. However, a detailed description of the mechanism of such a
measurement still lacks.} 
  properties of the non-abelian anyons are used to perform a quantum
measurement. 
But a perhaps more general applicable way to approach measurements, is
to look at \emph{interference experiments} with the quasi-particles,
both for non-abelian anyons, as suggested by \cite{preskill}, and
abelian anyons, \cite{ekert2}. Interference experiments with
non-abelian anyons are generalizations of the Aharonov-Bohm effect,
and have been studied in the past \cite{verlinde,lo}; however, these
papers focussed primarily on determining the cross-section via the
initial probability distribution, and neglected the \emph{change} in
the quantum state of the target particle. For topological quantum
computation, the quantum state of the target particle, i.e., the qubit
one likes to measure, obviously is important. 

Motivated by these considerations, we have analyzed certain types of
idealized interference experiments which one could perform with
non-abelian anyons, thereby taking into account the changes in the
quantum states of the quasi-particles. This leads to a critical study
of the non-abelian 
generalization of the Aharonov-Bohm effect and how it
could be measured by a Mach-Zender type device . It turns out that
there are essentially different possibilities to perform the repeated 
experiment leading to very different interference patterns. 

The paper is organized as follows, after a brief introduction to
non-abelian anyons and the braid group we discuss in section
\ref{sec:MZsetup} the 
general setup of experiments with the Mach-Zender interferometer. In
section \ref{sec:experiments} we use this setup to analyze the
different classes of 
anyonic experiments that could be performed and show to what kinds of
interference patterns they may lead. We distinguish the so-called {\em
one-to-one, many-to-one} and {\em many-to-many} experiments. The
observable patterns may be very distinct, because after a sufficient
number of repetitions the system will get locked into a particular
eigenstate or subspace of some well determined operator which depends
on the type of experiment. The proof of this locking is relegated to the
appendix. The probabilities of finding certain patterns is also
calculated. At the end of the paper we have summarized the results for
the various classes in a comprehensive way in Table
\ref{table:summary}.

\subsection{Non-abelian anyons and the braid group}
If we physically move two particles around each other and return both
to their original position, this corresponds to a closed path in the
two-particle configuration space.  In three or more dimensions this
closed loop will be contractible to a point, which means that the
monodromy operator describing the net effect on the two-particle
Hilbert-space will just be the identity operator. If we take two
identical (indistinguishable) particles one may also consider the
exchange operator defined as the square root of the monodromy
operator, which consequently can only have two distinct eigenvalues
namely plus or minus one corresponding to bosons and fermions
respectively.  It is well known that in two dimensions the situation
is fundamentally different; the loop is no longer contractible so that
the (unitary) monodromy operator can generate an arbitrary
phase. Similarly the exchange operator $R$ may yield {\em any} phase
on a state with two identical particles which are therefore denoted as
{\em anyons}. For such particles the anti-clockwise exchange operator
$\R$ and the clockwise exchange $\R^{-1}$ have to be distinguished.

On an algebraic level this leads to a considerable generalization
because it allows also the possibility of nontrivial braidings between
different types of particles. In short, the simple representation
theory of the {\em symmetric group} $S_n$ (of permutations) has to be
replaced by the representation theory of the {\em braid group} $B_n$
which is in principle an infinite group. The one dimensional unitary
representations of $B_n$ are labeled by an arbitrary phase, but its
higher dimensional unitary irreducible representations are far more
complicated.  In general the n-particle Hilbert space can be
decomposed in subspaces that will carry irreducible representations of
this braid group $B_n$ and if such representations turn out to be
higher dimensional we call the particles involved, {\em non-abelian
anyons}.

\begin{figure}[tbh]
\bc\parbox{9cm}{\narrowtext
\includegraphics{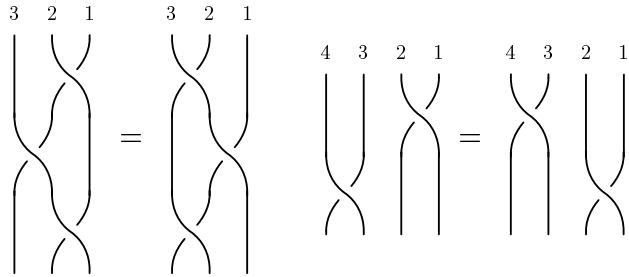}
\caption{A pictorial representation of the Yang-Baxter equations
$\tau_{i}\tau_{i+1}\tau_{i}=\tau_{i+1}\tau_{i}\tau_{i+1}$ and
$\tau_{i}\tau_j=\tau_j\tau_i$ with $|i-j|\ge2$ for $i=1$ and $j=3$;
the similarity with braids of strings of ropes is crystal-clear.} 
\label{fig:YBE}
\widetext}\ec
\end{figure}

The braid group $B_n$ is generated by $n-1$ elements $\tau_i$ and
their inverses. The $\tau_i$ obey the Yang--Baxter equations, also
known as the braid relations, which are: \bq\label{eq:YBE1}
\tau_{i}\tau_{i+1}\tau_{i}=\tau_{i+1}\tau_{i}\tau_{i+1}\quad\quad
i=1,\ldots,n-2 \eq \bq \tau_{i}\tau_{j}=\tau_{j}\tau_{i}\quad\quad
|i-j|\ge 2 \label{eq:YBE2}\eq 
The correspondence between the Yang--Baxter equations and braids of
strings of rope 
is best described pictorially, as in Fig. \ref{fig:YBE}.

The way in which non-abelian anyons are usually described, is by
endowing the particles with some internal degrees of freedom, the non
trivial braid statistics of these particles can than be implemented by
coupling these new internal degrees of freedom to a non-abelian
Chern-Simons (or statistical) gauge field which mediates the
appropriate type of topological interactions.

\begin{figure}[bth]
\bc\parbox{9cm}{\narrowtext
\includegraphics{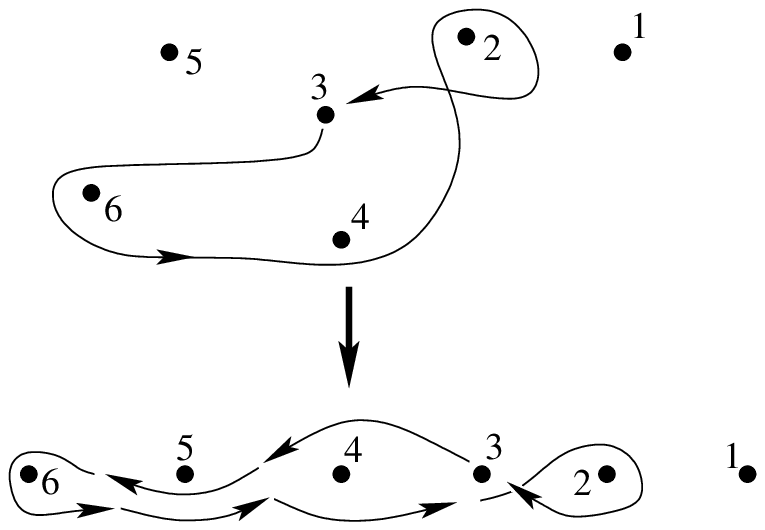}\\
\caption{An example on ordering a system with anyons and classifying
braids. We assign a number to all six particles and map them on a
(horizontal) line. The shown braid, which moves particle 3 around some
of the other particles and back to its original position, can now be
unambiguously decomposed in adjacent particle exchanges; read from
right to left this particular braid is given by:
$\R_2^{-2}\R_3\R_4\R_5^2\R_4^{-1}\R_3$.} 
\label{fig:braidorder}
\widetext}\ec
\end{figure}

As different braidings need not commute, it becomes important to be able 
to distinguish between them. A convenient way to accomplish this is to
order the system 
in the following way: map all particles in the two-dimensional plane
on some fixed 
line and number the particles (we will take the line to be horizontal
and we number r from right to left, starting with one).
If particles $i$ and $i+1$ pass each other on the virtual line, we
have to apply the appropriate exchange operator: 
$\R_i$ if the exchange is anti-clockwise, $\RI_i$ if clockwise (not
that after the exchange the original particle $i$ now is particle
$i+1$ and vice versa). 
Every braid can be decomposed as a product of these braids of adjacent
particles, $\R_i$ and $\RI_i$; also known as the braid operators,
$\R_i$ and $\RI_i$ clearly satisfy the braid relations
(\ref{eq:YBE1}), (\ref{eq:YBE2}). An example is shown in
Fig. \ref{fig:braidorder}. 

Associated with each particle is an internal space, and identical
particles carry identical internal spaces. Typically, and this
probably is the main feature of non-abelian anyons, the internal
states of different particles may become \emph{entangled} through
braiding: this interaction of the non-abelian anyons is of a
topological nature, independent of the distance between the anyons,
and it is a global property of the system, unperturbable by local
interactions. The total system's, multi-particle, possibly entangled,
internal state denoted by $\ket{\psi}$ lies in the tensor product of
the internal spaces of the individual particles. In such tensor
products we use the same ordering of particles as introduced above, in
that particles which are left (right) on the virtual line should
appear left (right) in the tensor product. For instance, imagine a
system with one particle of type $A$ (with internal space $V^A$) and
two particles of type $B$ (with internal spaces $V^B$) which lie left
of particle $A$. The three-particle internal state $\ket{\psi}$ is
then an element of $V^B\otimes V^B\otimes V^A$. 

The exchange caused by braid operators is made explicitly in the
tensor product as well, i.e.: 
$\R, \RI: V^B\otimes V^A\to V^A\otimes V^B$. We can think of $\R$ and
$\RI$ as combination of some matrix and an operator $\sigma$, where
$\sigma$ only swaps the two vector spaces. The matrix components of
$\R$ and $\RI$ depend only on the types of the two particles that is
operated upon. The square of $\R$, the monodromy operator $\R^2$,
obviously does no net swapping and can be regarded as a unitary
matrix, of which the explicit matrix-components depend on the
particles on which it operates. 

For the work we are about to present the knowledge of the
braidmatrices is assumed a priori. It is clearly of great importance
to understand the way one gets to this knowledge from an underlying
algebraic structure. For two-dimensional physics there turns out to be
an important generalization of ordinary group theory -- the theory of
quantum groups and Hopf algebra's -- which provides the natural
descriptions of non-abelian anyons, in that it generates natural
representations of the braid operator $R$ on the tensor product of its
representations. For the Fractional Quantum Hall Effect for example,
this programme has been carried out quite explicitly \cite{fqhe1,fqhe2}. This
has evolved 
into a rapidly expanding field of research on its own, we will however
not make explicit use of this machinery in the following.

\section{The setup of experiments with the Mach-Zender
interferometer}\label{sec:MZsetup} 
We introduce the setup of our thought experiments using the
Mach-Zender interferometer in this section. We will first treat an
ordinary interference experiment, then move to the Aharonov-Bohm
experiment, and generalize it further to the non-abelian anyon
experiment. We will see that for non-abelian anyons, there is a
nontrivial difference between the interference pattern and the
probability distribution.

\subsection{Ordinary interference}
We are interested in describing interference experiments with
non-abelian anyons, which basically means that we investigate
non-abelian generalizations of the Aharonov-Bohm effect. We should
realize that it is not the topological interactions leading to the
well known (phase) factors that cause the interference, for these
interactions will only be able to modify existing interference
patterns. So, the starting point is a basic device that produces
ordinary interference effects, as such a device, we may for example
choose the Mach-Zender interferometer. We will introduce some
terminology that we like to use in the following sections without
further reference.

\begin{figure}[!p]
\bc\parbox{14cm}{\mediumtext
\includegraphics{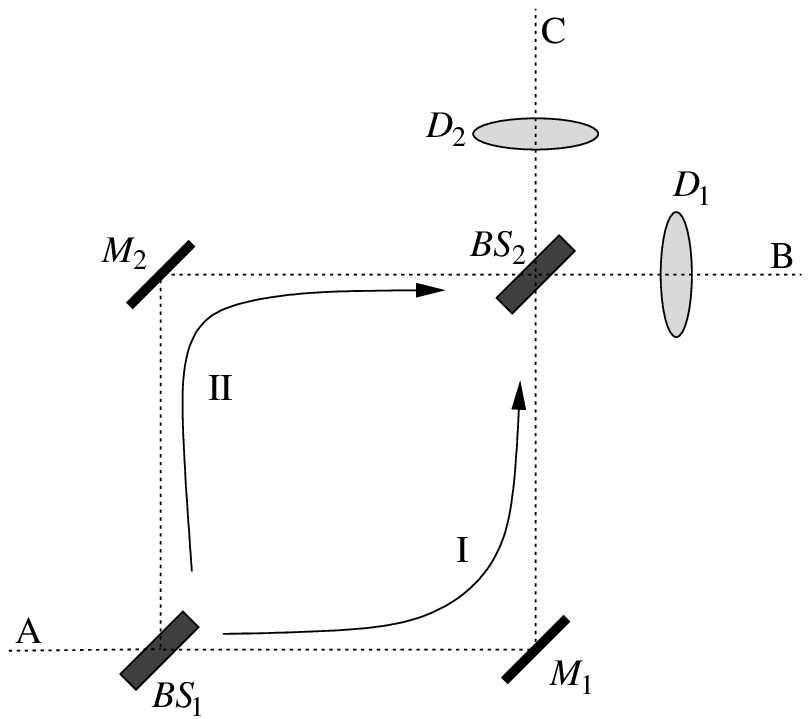}\\
\caption{The setup of the Mach-Zender interferometer for ordinary
interference experiments. At point A a particle may be injected in the
device, where it will encounter beam splitter $BS_1$. It may traverse
two different paths (thereby passing either mirror $M_1$ or $M_2$):
the counterclockwise path $\text{I}$ and pick up the phase factor
$e^{i\th_{\text{I}}}$, or the clockwise path $\text{II}$ and acquire
$e^{i\th_{\text{II}}}$. The particle is not restricted to one path
exclusively: it is allowed to travel both paths at the same time. Beam
splitter $BS_2$ unites both paths, which now interfere with each
other, and the particle emerges at one of two possible exits, B or C,
which will be observed by a detector, $D_1$ or $D_2$.}
\label{fig:MZ_OI}
\widetext}\ec
\end{figure}

The Mach-Zender interferometer, as depicted in Fig.~\ref{fig:MZ_OI},
is a two dimensional device, built of two lossless beam
splitters%
\footnote{We make use of the classical terminology of `splitting a
beam', but this should not distract us from the quantum nature of the
experiment: one particle at a time may propagate through the apparatus
and it is the two components of a single-particle wave packet that get
split by the beam splitter. When probed, by a detector like $D_1$ or
$D_2$, one would find there is but one particle and that it followed
one path, not both.}  $BS_1$ and $BS_2$, two 100\% reflecting mirrors
$M_1$ and $M_2$, and two detectors $D_1$ and $D_2$.  A
particle\footnote{We use `particle' in a rather abstract sense; it may
be any kind of particle as long as it exhibits both wave and particle
behavior. One may, for instance, think of electrons, or photons, but
in the present context also of quasi-particles with non-abelian
braiding properties.} enters the apparatus at point A on the left,
where it encounters a beam splitter $BS_1$. The beam splitter will
direct the particle  along the counterclockwise path $\text{I}$
and/or the clockwise path $\text{II}$. Both paths include a 100\%
reflecting mirror, $M_1$ or $M_2$, and both paths end up at beam
splitter $BS_2$. The particle may now emerge from the apparatus at
either point $B$ and be registered by detector $D_1$ or at point $C$
and be observed by detector $D_2$.

\begin{figure}[!p]
\bc\parbox{14cm}{\mediumtext
\includegraphics{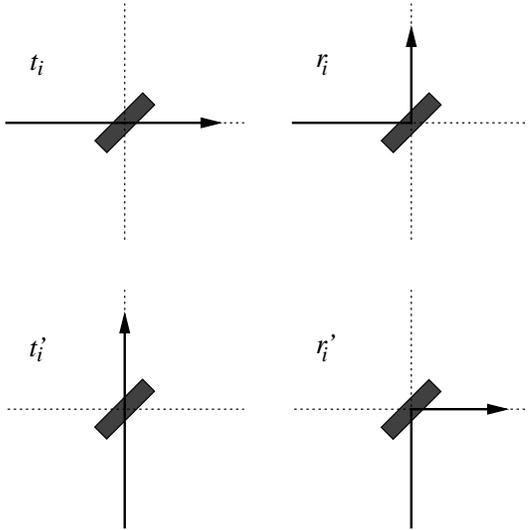}\\
\caption{The coefficients $t_i$, $r_i$, $t'_i$, $r'_i$ determine the
lossless beam splitter $BS_i$ (where $i=1,2$); $t_i$ indicates
transmission for incoming beams from the left, $r_i$ stands for
reflection when incident from the left, $t'_i$ and $r'_i$ denote
transmission and reflection when incident from below.}
\label{fig:BSi_coef}
\widetext}\ec
\end{figure}

Ordinary interference is caused by the difference in relative phase of
particles traveling along path $\text{I}$ or path $\text{II}$, which
is due to a difference in path length. We assume that we can adjust
the apparatus in such a way that we can tune the difference in path
length between paths $\text{I}$ and $\text{II}$ with high
precision. We also have to include the transmission and reflection
coefficients $t_i,r_i,t'_i,r'_i$ of the beam splitters $BS_i$, see
Fig.~\ref{fig:BSi_coef}, because these complex factors also affect the
relative phase. Since we assume the beam splitters to be perfect, its
coefficients form a unitary matrix and we have that
\cite{zeilinger:1981,silverman:mystery}: \bq r_i^*=r_i',\quad\quad
t_i^*=-t'_i,\quad\quad |r_i|+|t_i|^2=1.\label{eq:rt_unitary} \eq

We will not explicitly need the lengths of paths $\text{I}$ and
$\text{II}$, only the relative phases that the particle will
acquire when traversing path $\text{I}$ or $\text{II}$. We will
indicate the associated phase factors by $e^{i\th_{\text{I}}}$ and
$e^{i\th_{\text{II}}}$. The amplitude $A[D_1]$ of the particle's wave
packet component at detector $D_1$ is the sum of the contributions of
both paths with the appropriate factors: \bq
A[D_1]=t_1r'_2e^{i\th_{\text{I}}}+r_1t_2 e^{i\th_{\text{II}}}, \eq and
for $A[D_2]$: \bq A[D_2]=t_1t'_2e^{i\th_{\text{I}}}+r_1r_2
e^{i\th_{\text{II}}}.  \eq The probabilities $P[D_1]$ and $P[D_2]$
that indicate the probability for the particle to be detected at $D_1$
or $D_2$ are the absolute squares of the amplitudes $A[D_1]$ and
$A[D_2]$: \bq
\label{eq:Pordint1a}P[D_1]=|t_1r'_2|^2+|r_1t_2|^2+2
\Re\left(t_1r'_2r^*_1t^*_2e^{i(\th_{\text{I}}-\th_{\text{II}})}\right), 
\eq
\bqa
\label{eq:Pordint2a}P[D_2]&=&|t_1t'_2|^2+|r_1r_2|^2+2
\Re\left(t_1t'_2r^*_1r^*_2e^{i(\th_{\text{I}}-\th_{\text{II}})}\right)\\ 
\label{eq:Pordint3a}&=&|t_1t'_2|^2+|r_1r_2|^2-2
\Re\left(t_1t^*_2r^*_1r'_2e^{i(\th_{\text{I}}-\th_{\text{II}})}\right), 
\eqa where we used (\ref{eq:rt_unitary}) to go from
(\ref{eq:Pordint2a}) to (\ref{eq:Pordint3a}). Adding
eqs.~(\ref{eq:Pordint1a}) and (\ref{eq:Pordint3a}) confirms that the
total probability is of course preserved: \bq P[D_1]+P[D_2]=1.
\eq In a real experimental setup, the particles that are directed at
the apparatus will not all acquire precisely the same phase factors
$e^{i\th_{\text{I}}}$ or $e^{i\th_{\text{II}}}$, for instance because
of a slight difference in the momenta of the incoming
particles. Therefore, we introduce a variable $q$ and a density
function $\rho(q)$ to incorporate all such factors that influence the
relative phase factors (which are due to the experimental setup). When
calculating probabilities, we have to integrate over the, now $q$
dependent, phase factors $e^{i\th_{\text{I}}(q)}$,
$e^{i\th_{\text{II}}(q)}$: \bq \int
e^{i[\th_{\text{I}}(q)-\th_{\text{II}}(q)]}\rho(q)dq\equiv Qe^{i\th},
\eq where $Q$ is real and between zero and one,%
\footnote{If $Q$ is zero no interference is observed, because the
phase factors average out to zero; this is a common aspect in
\emph{every} interference experiment: the density distribution
$\rho(q)$ needs to be narrow enough to keep observable interference.}  and
$\th$ is some sort of average phase difference. After integration over
$q$, the probabilities of eqs.~(\ref{eq:Pordint1a}) and
(\ref{eq:Pordint3a}) then become: \bq
\label{eq:Pordint1b}P[D_1]=|t_1r'_2|^2+|r_1t_2|^2+2Q
\Re\left(t_1r'_2r^*_1t^*_2e^{i\th}\right), 
\eq
\bq
\label{eq:Pordint3b}P[D_2]=|t_1t'_2|^2+|r_1r_2|^2-2Q
\Re\left(t_1r'_2r^*_1t^*_2e^{i\th}\right). 
\eq We will use $e^{i\th_{\text{I}}}$ and $e^{i\th_{\text{II}}}$ in
our notation for amplitudes, and $Q$ and $e^{i\th}$ for probabilities,
without further mentioning the implied $q$-integration.

So far, we have only talked about probabilities for single particles
to be injected and observed. Obviously, we want to repeat many of such
single particle `runs' and call the result the interference
pattern. In other words, the interference pattern $I[D_1]$ is the
number of times $\#[D_1]$ that a particle is injected in the apparatus
and observed by detector $D_1$ divided by the total number $n$ of
injected particles, and likewise for $I[D_2]$: \bq
I[D_1]=\frac{\#[D_1]}{n},\quad\quad
I[D_2]=\frac{\#[D_2]}{n},\label{eq:def_oi_ip} \eq \bq
\#[D_1]+\#[D_2]=n.  \eq If we repeat the experiment for a large number
of particles, i.e. perform many runs under the same conditions, we
expect \emph{by definition} that the observed interference patterns
become equal to the probability distributions: \bq
\lim_{n\to\infty}I[D_1]=P[D_1],\quad\quad\lim_{n\to\infty}I[D_2]=P[D_2].
\eq

On several occasions we will give examples, where for definiteness we have
chosen the following values for the coefficients and factors that
completely determine the Mach-Zender apparatus:
\[
r_j=r'_j=\frac{1}{\sqrt{2}},\quad\quad
t_j=-t'_j=i\frac{1}{\sqrt{2}},\quad\quad (j=1,2)
\]
\bq Q=1,\quad\quad \th=\arccos(\case{4}{5}).\label{eq:coef_example}
\eq These values were chosen such that, when substituted in
eqs.~(\ref{eq:Pordint1b}) and (\ref{eq:Pordint3b}), the probabilities
$P[D_1]$ and $P[D_2]$ become nice, but still representative, numbers:
\bq P[D_1]=\case{9}{10}\quad\quad
P[D_2]=\case{1}{10}\label{eq:OI_example}.  \eq So, in this particular
example, it is by far more likely to detect the emerging particle at
detector $D_1$; only 10\% of the incident particles is observed at
detector $D_2$.

\subsection{The Aharonov-Bohm effect: the topological phase factor}
We will now discuss the Aharonov-Bohm version of the Mach-Zender
interferometer. The Aharonov-Bohm effect is the most logical step in
going from ordinary interference experiments to interference
experiments with non-abelian anyons, as it is both a well understood
extension of ordinary interference and at the same time features a
topological aspect.

\begin{figure}[!tbh]
\bc\parbox{9cm}{\narrowtext
\includegraphics{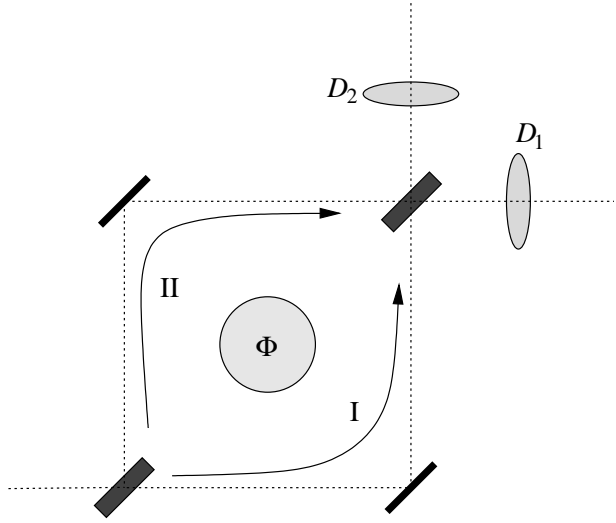}\\
\caption{The Aharonov-Bohm effect in the setup of the Mach-Zender
interferometer: the presence of $\Phi$ contributes to additional
topological phase factors. Following path $\text{I}$ gives rise to the
factor $e^{i\l_{\text{I}}}$, and traversing path $\text{II}$ yields
$e^{i\l_{\text{II}}}$. The incoming particle still emerges at either
detector $D_1$ or $D_2$, but the probabilities for either exit have
been altered. The interference pattern now depends explicitly on the
difference of the relative phases
$e^{i\l}=e^{i(\l_{\text{I}}-\l_{\text{II}})}$.}
\label{fig:MZ_AB}
\widetext}\ec
\end{figure}

The Mach-Zender apparatus is extended with a physical object $\Phi$
that we put at its center, as depicted in Fig.~\ref{fig:MZ_AB}, and we
assume that this object has a topological interaction with the
injected particles.%
\footnote{In case of the famous Aharonov-Bohm effect experiment where
electrons are the incoming particles, $\Phi$ represents a magnetic
flux, and the electron picks up the phase factor $e^{iae\Phi}$ ($a$ is
some constant and $e$ the charge of the electron) encircling the flux
once.}
The incoming particle now acquires an additional phase factor which
\emph{only} depends on the topology of the path (where from a
topological point of view $\Phi$ is a puncture of the two-dimensional
plane). Paths $\text{I}$ and $\text{II}$ are topologically distinct
and contribute the additional phase factors $e^{i\l_{\text{I}}}$ and
$e^{i\l_{\text{II}}}$. The amplitude $A_{\text{AB}}[D_1]$ of the
particle's wave packet component at detector $D_1$ for the
Aharonov-Bohm effect obviously includes these phase factors: \bq
A_{\text{AB}}[D_1]=t_1r'_2e^{i\th_{\text{I}}}e^{i\l_{\text{I}}}+r_1t_2
e^{i\th_{\text{II}}}e^{i\l_{\text{II}}}, \eq and likewise for the
amplitude $A_{\text{AB}}[D_2]$ of the component at detector $D_2$.

The probabilities $P_{\text{AB}}[D_i]$ to observe a particle at either
detector depend on the (gauge invariant) phase difference
$e^{i(\l_{\text{I}}-\l_{\text{II}})}\equiv e^{i\l}$, which is the
topological phase factor that a particle picks up when it circumvents
$\Phi$ in a counterclockwise way. The probabilities
$P_{\text{AB}}[D_i]$ are: \bq
\label{eq:Pab1}P_{\text{AB}}[D_1]=|t_1r'_2|^2+|r_1t_2|^2+2Q
\Re\left(t_1r'_2r^*_1t^*_2e^{i\th}e^{i\l}\right), 
\eq \bq
\label{eq:Pab2}P_{\text{AB}}[D_2]=|t_1t'_2|^2+|r_1r_2|^2-2Q
\Re\left(t_1r'_2r^*_1t^*_2e^{i\th}e^{i\l}\right). 
\eq As expected, the particle still emerges from the apparatus
with unit probability: \bq P_{\text{AB}}[D_1]+P_{\text{AB}}[D_2]=1.
\eq We will use the above expressions for $P_{\text{AB}}[D_i]$ later
on, where we also want to explicitly include the dependence on
$e^{i\l}$. For this purpose we introduce $P_{e^{i\l}}[D_i]$, which is
defined as: \bq P_{e^{i\l}}[D_i]\equiv
P_{\text{AB}}[D_i].\label{eq:Plambda} \eq

The single particle run can be repeated for many particles, which
gives rise to the interference patterns $I_{\text{AB}}[D_i]$ similar
to those of eq.~(\ref{eq:def_oi_ip}). And again, when the number $n$ of
particles grows, the interference patterns approach the probability
distributions: \bq
\lim_{n\to\infty}I_{\text{AB}}[D_1]=P_{\text{AB}}[D_1]=P_{e^{i\l}}[D_1],
\eq \bq
\lim_{n\to\infty}I_{\text{AB}}[D_2]=P_{\text{AB}}[D_2]=P_{e^{i\l}}[D_2].
\eq

This is about all there is to say about the Aharonov-Bohm interference
experiment apart from an example that demonstrates how $e^{i\l}$
changes the interference pattern. Let us substitute $e^{i\l}=-1$ and
the values from (\ref{eq:coef_example}) into (\ref{eq:Pab1}) and
(\ref{eq:Pab2}); this yields: \bq
P_{\text{AB}}[D_1]=\case{1}{10}\quad\quad
P_{\text{AB}}[D_2]=\case{9}{10}, \eq where the particle is now with
90\% chance detected by detector $D_2$ in contrast with the values of
the probabilities in the example of the ordinary interference
experiment given by (\ref{eq:OI_example}).

\subsection{The setup with non-abelian anyons}
\begin{figure}
\bc\parbox{9cm}{\narrowtext
\includegraphics{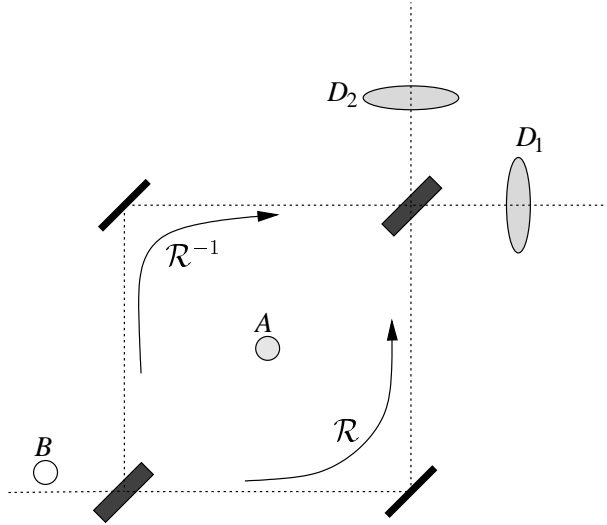}\\
\caption{The setup for interference experiments with non-abelian
anyons using the Mach-Zender interferometer. At the center of the
apparatus resides a non-abelian anyon, particle $A$. The non-abelian
anyon incident from the left is particle $B$. The non-abelian anyons
have a topological interaction; if particle $B$ traverses the
counterclockwise path this corresponds to the braid operator $\R$
acting on the two-particle internal state, the clockwise path yields
the inverse operation $\RI$. The probabilities to emerge at detector
$D_1$ and $D_2$ will explicitly depend on the expectation value of the
monodromy operator $\R^2$.}
\label{fig:MZ_NA}
\widetext}\ec
\end{figure}

The setup for - what up to now should be considered thought
experiments with non-abelian anyons using the Mach-Zender
interferometer, is very similar to the Aharonov-Bohm effect. The
incoming particle will be a non-abelian anyon and the object $\Phi$ at
the center of the apparatus will be replaced by another particle which
is {\em also} a non-abelian anyon. So, it is no longer a single
particle problem (one particle scattering of an invariant external
object $\Phi$), but a two particle problem (one particle scattering of
another one, both being non-abelian anyons). We will designate the
incident non-abelian anyon as particle $B$ and the non-abelian anyon
fixed at the apparatus' center as particle $A$. The idea of the
experiment remains the same as for the conventional interference
experiments: we inject the particle $B$, in the Mach-Zender
apparatus and detect it again with one of the two detectors $D_1$ and
$D_2$, also see Fig.~\ref{fig:MZ_NA}.

In calculating the amplitudes and probabilities for the non-abelian
anyon experiment we have to include the two-particle internal state
$\ket{\psi}$, which is an element of the tensor product of the
internal spaces $V^B$ and $V^A$ of particles $B$ and $A$: \bq
\ket{\psi}\in V^B\otimes V^A \eq. The topological phase factors in the
Aharonov-Bohm effect will be replaced by braid operators acting on the
state $\ket{\psi}$. If particle $B$ follows path $\text{I}$ within the
apparatus it picks up a counterclockwise braid on the two-particle
internal state: $\R\ket{\psi}$; if particle $B$ would traverse path
$\text{II}$ this would correspond to a clockwise braid:
$\RI\ket{\psi}$.

Let us write down the amplitudes $A_{\text{NA}}[D_i]$ for the
interference experiment with non-abelian anyons: \bq
A_{\text{NA}}[D_1]=t_1r'_2e^{i\th_{\text{I}}}\R\ket{\psi}+r_1t_2
e^{i\th_{\text{II}}}\RI\ket{\psi},
\label{eq:Ana1}\eq
\bq
A_{\text{NA}}[D_2]=t_1t'_2e^{i\th_{\text{I}}}\R\ket{\psi}+r_1r_2
e^{i\th_{\text{II}}}\RI\ket{\psi}. 
\label{eq:Ana2}\eq
In calculating the probability distributions $P_{\text{NA}}[D_i]$, we
use that the braid operators are unitary and $\ket{\psi}$ is
properly normalized: \bq
\label{eq:Pna1}P_{\text{NA}}[D_1]=|t_1r'_2|^2+|r_1t_2|^2+2Q
\Re\left(t_1r'_2r^*_1t^*_2e^{i\th}\bra{\psi}\R^2\ket{\psi}\right), 
\eq
\bq
\label{eq:Pna2}P_{\text{NA}}[D_2]=|t_1t'_2|^2+|r_1r_2|^2-2Q
\Re\left(t_1r'_2r^*_1t^*_2e^{i\th}\bra{\psi}\R^2\ket{\psi}\right). 
\eq Compared with the probabilities $P_{\text{AB}}[D_i]$ for the
Aharonov-Bohm effect, equations (\ref{eq:Pab1}) and (\ref{eq:Pab2}),
the topological phase factor $e^{i\l}$ has been replaced by the
expectation value $\bra{\psi}\R^2\ket{\psi}$ of the monodromy operator
$\R^2$ in (\ref{eq:Pna1}) and (\ref{eq:Pna2}).

We can re-write the probabilities $P_{\text{NA}}[D_i]$ when we use
that the monodromy operator is unitary and thus has eigenvalues of the
form $e^{i\l}$. If we introduce the projection operators $E_\l$ which
project onto the eigenspace of $e^{i\l}$, we have that: \bq
\R^2=\sum_\l e^{i\l}E_\l,\quad\quad
p_\l\equiv\bra{\psi}E_\l\ket{\psi},\label{eq:El1} \eq \bq
\bra{\psi}\R^2\ket{\psi}=\sum_\l p_\l e^{i\l},\quad\quad\sum_\l
E_\l=\openone,\quad\quad\sum_\l p_\l=1,\label{eq:El2} \eq where we
indicated the real-valued expectation value of $E_\l$ by $p_\l$. With
eq.~(\ref{eq:Plambda}) in mind, the probability distribution of
(\ref{eq:Pna1}) can now be cast in the following form: \bqa
P_{\text{NA}}[D_1]&=&|t_1r'_2|^2+|r_1t_2|^2+2Q
\Re\,\Bigl(t_1r'_2r^*_1t^*_2e^{i\th}\sum_\l 
p_\l e^{i\l}\Bigr)\nn\\ &=&\sum_\l
p_\l\Bigl[|t_1r'_2|^2+|r_1t_2|^2+2Q
\Re\left(t_1r'_2r^*_1t^*_2e^{i\th}e^{i\l}\right)\Bigr]\nn\\ 
&=&\sum_\l p_\l P_{e^{i\l}}[D_1]\label{eq:Pna1b}.  \eqa This equation
is nice because it gets rid of the
uninteresting coefficients $r_i$,$Q$, etc; it states
that the probability to observe particle $B$ at detector $D_1$ is a
sum over probabilities of the Aharonov-Bohm version of the experiment,
where the eigenvalues $e^{i\l}$ of the monodromy operator $\R^2$ play
the role of the topological phase factor. The weight $p_\l$ in the sum
is precisely the same weight as in the decomposition of (\ref{eq:El2})
of $\bra{\psi}\R^2\ket{\psi}$ into eigenvalues $e^{i\l}$. Of course,
there is a similar decomposition for $P_{\text{NA}}[D_2]$, so we can
write for $P_{\text{NA}}[D_i]$: \bq P_{\text{NA}}[D_i]=\sum_\l p_\l
P_{e^{i\l}}[D_i]\label{eq:Pna2b}, \eq and the total probability still
equals unity: \bq P_{\text{NA}}[D_1]+P_{\text{NA}}[D_2]=\sum_{j,\l}
p_\l P_{e^{i\l}}[D_j]=1.  \eq

Clearly, the values of the probabilities $P_{\text{NA}}[D_i]$ depend
on the two-particle internal state $\ket{\psi}$, via the weights
$p_\l$. Let us give an example to illustrate this. We plug the
coefficients of the Mach-Zender apparatus given by
(\ref{eq:coef_example}) into eq.~(\ref{eq:Pna2b}), and assume that
the monodromy operator $\R^2$ has two eigenvalues: $\!+1,-1$
(eq.~(\ref{eq:explicitR2}) shows a specific example of such a matrix
$\R^2$). In the following table, we have given $P_{\text{NA}}[D_i]$
for three internal states, 
each of which is characterized by the $p_\l$; we also included the
expectation value 
$\bra{\psi}\R^2\ket{\psi}$.\\

\begin{quasitable}
\bc\parbox{9cm}{\narrowtext
\begin{tabular}{ccccc}
\tableline\tableline
&&&\multicolumn{2}{c}{Probability distributions}\\
$\Big.$$p_{1}$&$p_{-1}$&$\bra{\psi}\R^2\ket{\psi}$&$P_{\text{NA}}[D_1]$& 
$P_{\text{NA}}[D_2]$ \\ 
\tableline
$\Big.$$\case{1}{2}$	&$\case{1}{2}$	&$0$	&$\frac{1}{2}$
&$\frac{1}{2}$\\ 
$\Big.$$1$		&$0$		&$1$	&$\frac{9}{10}$
&$\frac{1}{10}$\\ 
$\Big.$$0$		&$1$		&$-1$	&$\frac{1}{10}$
&$\frac{9}{10}$\\ 
\tableline\tableline
\end{tabular}\widetext}\ec
\end{quasitable}

It is worth mentioning that one is not
restricted to the Mach-Zender interferometer to make the decomposition
of the probabilities, like in expression (\ref{eq:Pna2b}). This is not
so strange, since any dependence on the Mach-Zender apparatus, i.e.,
the coefficients $r_i$,$Q$ etc., is absent in (\ref{eq:Pna2b}). For
instance, for double slit or plain scattering interference experiments
with non-abelian anyons one can write down a similar
decomposition. However, it is not conventional to do so; one usually
keeps the expectation value of $\R^2$ explicit in one's expressions.

\subsection{From probabilities to interference patterns}
The interference experiment with non-abelian anyons is obviously a
generalization of the Aharonov-Bohm experiment, by replacing the phase
factor $e^{i\l}$ by a unitary operator $\R^2$, which has eigenvalues
of the form $e^{i\l}$. But the picture of the interference experiment
with non-abelian anyons is not complete yet, and requires further
analysis. We have to consider the question of interference
patterns, which for the case of non-abelian
anyons we avoided so far.

\begin{figure}
\bc\parbox{9cm}{\narrowtext
\includegraphics{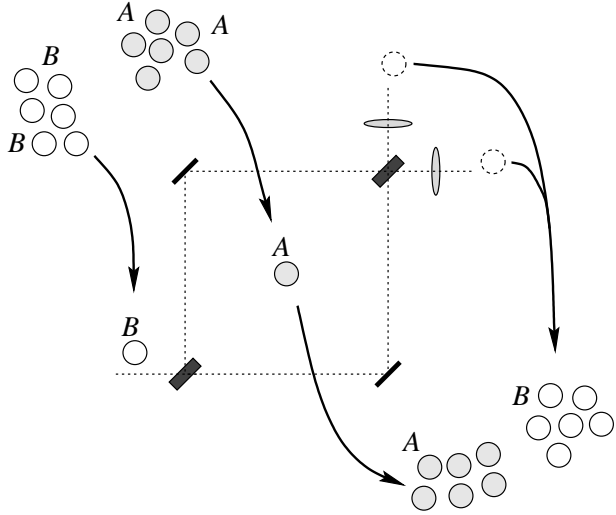}\\
\caption{The {\em many-to-many} experiment. To be able to measure the
initial probability distribution as the interference patterns requires
that exactly the same run be repeated often. This means that after
each observation of particle $B$ at either detector both particles $A$
and $B$ have to be replaced by new ones having the same initial internal
state. The old particles cannot be re-used because the detection of
particle $B$ at $D_i$ causes a change in the internal state. This way
to repeat the runs is called the many-to-many experiment as it
involves many particles $B$ and many particles $A$.}
\label{fig:MZ_mm}
\widetext}\ec
\end{figure}

What one wants is, to predict what one would observe in a realistic
experiment. The interference pattern is the number of particles the
detectors count if the experiment is repeated with many particles. For
the ordinary interference and the Aharonov-Bohm experiments, is was
perfectly legitimate to identify the interference patterns with the
probability distributions, since by definition these are equal when
the same experiment is repeated many times. Is this also true for the
experiment with non-abelian anyons? The answer to this question turns
out to be: not necessarily. The problem lies in the definition: in
order to be the same one needs to repeat exactly the same
experiment. For the non-abelian anyons experiment, this would imply
that for each consecutive run, \emph{both} particle $A$ and $B$ would
have to be discarded, to be replaced by new particles $A$ and
$B$, as is also shown in Fig.~\ref{fig:MZ_mm}.

Why do we need a new particle $A$ and the same particle $B$ each time,
why not use the old particles? The answer is, the two-particle
internal state $\ket{\psi}$ not only affects the probabilities to
observe particle $B$ at either detector, but is also altered itself by
the detection, becoming a new, different state $\ket{\psi'}$! It is
obvious that an experiment with the internal state $\ket{\psi}$ gives
another result than an experiment with internal state $\ket{\psi'}$;
so indeed, if we want to find the calculated probabilities as the
observed interference pattern, we need a large supply of particles $A$
and $B$, all with the same internal states $\ket{\psi}$. Then we have
that: \bq \lim_{n\to\infty}
I^{\text{m-m}}_{\text{NA}}[D_i]=\lim_{n\to\infty}
\frac{\#[D_i]}{n}=P_{\text{NA}}[D_i], \eq where
$I^{\text{m-m}}_{\text{NA}}[D_i]$ indicate the interference patterns
for the experiment with non-abelian anyons that is repeated with many
particles $B$ directed to many particles $A$. We will refer to this
type of experiment as a \emph{many-to-many} (m-m) experiment.

However, we may decide to repeat the experiment in a different way,
i.e., not using a new $A$ and $B$ each time, but then we will have to
include the change in the internal state. The change in the
two-particle internal state from $\ket{\psi}$ to $\ket{\psi'}$ itself
is also worth examining, and observable as well; if we want to observe
the effect of this change of $\ket{\psi}$, it is \emph{necessary} that
we repeat the experiment while keeping at least one of the two old
particles $A$ and $B$. We will discuss two such ways to repeat the
experiment.  First we will describe an experiment in which we keep using the
same particle $A$ and particle $B$ for all consecutive runs, we
call this the \emph{one-to-one} experiment. The second way to
repeat the experiment, is to keep only particle $A$ and replace each
particle $B$ by a new one for each run; we will call this type of
experiment \emph{many-to-one}.%
\footnote{One may wonder if there are more ways to repeat an
interference experiment with non-abelian anyons. There are endless
ways to do so, but these are all hybrids of the many-to-many,
one-to-one and many-to-one experiments, and introduce no new
concepts. On might be tempted to say a one-to-many experiment is
missing; however, the setup is symmetrical under the exchange of
particle $A$ and $B$, and a one-to-many experiment would yield the
same results as the many-to-one experiment does.}

The reason that the two-particle internal state $\ket{\psi}$ gets
changed, is because the detection of particle $B$ is a quantum
mechanical measurement; there are two components of the two-particle
wave packet, one associated with detector $D_1$, the other with $D_2$,
and the measurement process projects out one of these components. The
amplitudes $A_{\text{NA}}[D_i]$, (\ref{eq:Ana1}) and (\ref{eq:Ana2}),
describe these components: the $D_1$ component is
$t_1r'_2e^{i\th_{\text{I}}}\R\ket{\psi}+r_1t_2
e^{i\th_{\text{II}}}\RI\ket{\psi}$ and the $D_2$ component is
$t_1t'_2e^{i\th_{\text{I}}}\R\ket{\psi}+r_1r_2
e^{i\th_{\text{II}}}\RI\ket{\psi}$. If particle $B$ is detected by
$D_1$ the two-particle internal state $\ket{\psi}$ changes to
$\ket{\psi'}$, which is the normalized $D_1$ component: \bq
\ket{\psi'}=\frac{1}{\sqrt{K}}\left(t_1r'_2e^{i\th_{\text{I}}}
\R\ket{\psi}+r_1t_2 
e^{i\th_{\text{II}}}\RI\ket{\psi}\right),\label{eq:componentprojection}
\eq where the real factor $K$ is the normalization constant. There is
a similar expression for the case particle $B$ is detected by detector
$D_2$.

A priori, we may not assume that we know the precise values of $t_i$,
$r_i$, $r'_i$, $e^{i\th_{\text{I}}}$, $e^{i\th_{\text{II}}}$ in the
above expression (\ref{eq:componentprojection}); it is also doubtful
whether we know $\ket{\psi}$ beforehand, but even if we do, we know
little of the new state $\ket{\psi'}$. The best we can say is that
$\ket{\psi'}$ will be an entangled state, i.e., it cannot be
factorized  as a simple tensor product $\ket{x}\otimes\ket{y}$, $\ket{x}\in
V^A$, $\ket{y}\in V^B$; this is because $\ket{\psi'}$ is, see
eq.~(\ref{eq:componentprojection}), some linear combination of $\R$
and $\RI$ acting on $\ket{\psi}$, and for non-trivial $\R$ and $\RI$
this action usually results in entanglement, independent of whether
the initial state $\ket{\psi}$ was entangled. However, as we will see
in the one-to-one and the many-to-one experiment, it turns out that we
can work with this expression for $\ket{\psi'}$ and eventually, after
observing many incident particles at the detectors, we can even
determine $\ket{\psi'}$ with precision.

\section{The inequivalent classes of interference
experiments}\label{sec:experiments} 
In this section we want to consider the various possibilities for the
interference experiments in detail. We start with the one-to-one
experiment followed by the many-to-one experiment (we already
discussed the many-to-many experiment in the previous section). We
conclude by 
summarizing the different results in a Table \ref{table:summary} .

\subsection{The one-to-one experiment}
\begin{figure}
\bc\parbox{9cm}{\narrowtext
\includegraphics{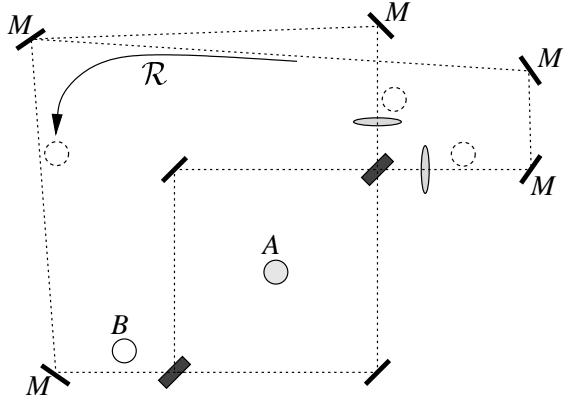}\\
\caption{The setup for the {\em one-to-one} experiment, which uses only one
particle $B$ and one particle $A$. Additional (intelligent) mirrors
$M$ have been placed around the Mach-Zender device to route particle
$B$ back to its original position after it has been detected by $D_1$
or $D_2$, where it will be injected again in the apparatus. As
particle $B$ is returned, it passes particle $A$ in a counterclockwise
fashion, which means the braid operator $\R$ should act on the
two-particle internal state. Particle $A$ remains all the time
unperturbed at the center of the apparatus. With each detection of
particle $B$, the two-particle internal state may change.}
\label{fig:MZ_oo}
\widetext}\ec
\end{figure}

The one-to-one experiment requires only one
particle $A$ and one particle $B$. What we will show  in this
subsection, is that after repeating the one-to-one experiment a
sufficient number of times, the system will get locked into an
eigenstate of the monodromy operator $\R^2$ with a probability
determined by the initial state.  We inject particle $B$ at the left
of the Mach-Zender apparatus, with particle $A$ at its center, and
detect particle $B$ at detector $D_1$ or $D_2$, and we repeat this
procedure using the same particles $B$ and $A$. This of course
requires, that we
route particle $B$ back to its original position - the
entrance point at the left of the apparatus. We assume that we can extend the
Mach-Zender device with some additional mirrors that take care of
this, as depicted in Fig.~\ref{fig:MZ_oo}.%
\footnote{These mirrors should be somewhat smart, in the sense that
they may adjust their position based on whether particle $B$ emerged
at $D_1$ or $D_2$, to make sure that particle $B$ is returned at the
same position independent of the detector at which it was observed at
the previous run.}%
\footnote{Particle $B$ might need some preparation as to become a wave
packet suitable to `split' at the beam splitter, for instance it may
need to be accelerated. Devices, needed for such purposes, can be
placed to the left of the Mach-Zender apparatus, but we will not
bother ourselves with such devices as they do not contribute to the
essence of the problem.}  As we move non-abelian anyons around, we
should be aware that this may include some action of the braid
operators; as we move particle $B$ back, we exchange it with particle
$A$ in a counterclockwise way, which means that an additional braid
operation $\R$ acts on the two-particle internal state $\ket{\psi'}$.

Let us take a closer look at the one-to-one experiment, paying special
attention to the internal state. We start with an initial two-particle
internal state $\ket{\psi}\equiv\ket{\psi_0}$, and insert particle $B$
in the apparatus. We then observe particle $B$ at one of the detectors
$D_i$. The probability to emerge at detector $D_i$ is of course given
by (\ref{eq:Pna1}):
$P_{\text{NA}}[D_i]=P_{\text{NA}}[D_i]_{\ket{\psi_0}}$, to show
explicitly the dependence on the internal state $\ket{\psi_0}$. Since
the probability for each detector will in general be non-zero,
there are two possible outcomes for the first run: particle $B$ is
detected at $D_1$ or at $D_2$. If it is at detector $D_1$ and we
return particle $B$ to its initial position, the new two-particle
internal state $\ket{\psi_1}$ (we will use $\ket{\psi_j}$ to indicate
the internal state after the $j$th run) will be: \bqa 
\ket{\psi_1}&=&\R\frac{1}{\sqrt{K}}\left(t_1r'_2e^{i\th_{\text{I}}}
\R\ket{\psi}+r_1t_2  
e^{i\th_{\text{II}}}\RI\ket{\psi}\right)\nn\\
&=&\frac{1}{\sqrt{K}}\left(t_1r'_2e^{i\th_{\text{I}}}\R^2+r_1t_2
e^{i\th_{\text{II}}}\openone\right)\ket{\psi_0}.\label{eq:stateevolution1}
\eqa If we would find particle $B$ at detector $D_2$, the new state
$\ket{\psi_1}$ would be: \bq
\ket{\psi_1}=\frac{1}{\sqrt{K'}}\left(t_1t'_2e^{i\th_{\text{I}}}\R^2+r_1r_2
e^{i\th_{\text{II}}}\openone\right)\ket{\psi_0},\label{eq:stateevolution2}
\eq where $K'$ is a normalization constant that will in general be
different from $K$.

We can easily continue this for a second run. The probability
to observe particle $B$ at detector $D_i$ at the second run is given
by $P_{\text{NA}}[D_i]_{\ket{\psi_1}}$, where we note that
$\ket{\psi_1}$, albeit implicitly, depends on the outcome of the first
run. Of course, there are two possible outcomes for the second run,
and the resulting internal state $\ket{\psi_2}$ will depend on the
outcomes of the first and second run. We could follow this procedure
for a large number of runs, say $n$. However, this means we would have
to work with an internal state $\ket{\psi_n}$ depending on the
outcomes of all $n$ runs, meaning there would be $2^n$ possible
outcomes to consider!  
Although this would probably lead to the correct
answer, the number $2^n$ grows so incredibly fast, that this way is
not advisable, and indeed we will follow another path to the answer.%
\footnote{Actually, the first thing we did was a numerical
simulation. We did not consider all $2^n$ outcomes, but tried to
simulate the `real' thought experiment: guided by probabilities we
followed branches in the `tree' of all $2^n$ outcomes. These numerical
simulations lead us straight to the results we are presenting here.} 

We are basically interested in two things after we repeated the
experiment for $n$ runs: firstly the interference pattern, i.e., how
many times $\#[D_i]$ out of the total $n$ we observed the incident
particle at detector $D_i$, and secondly, what we can say about the
internal state $\ket{\psi_n}$ given the interference pattern.

The interference pattern has something to do with the probabilities
$P_{\text{NA}}[D_i]_{\ket{\psi_j}}$, $j=1,2,\ldots,n$. Since
$\ket{\psi_1}$ is different from $\ket{\psi_0}$, which is in the
general  case 
true and independent of the outcome of the first run, we can write down
the following inequality concerning their probability distributions:
\bq P_{\text{NA}}[D_i]_{\ket{\psi_1}}\ne
P_{\text{NA}}[D_i]_{\ket{\psi_0}}.\label{eq:Pinequality} \eq One can
wonder if such an inequality holds also for the $j$th and $(j+1)$th
run. This would mean that the probabilities keep fluctuating from run
to run and the resulting interference pattern would become random, and
thereby also independent on the initial internal state
$\ket{\psi_0}$. However, it turns out that this is not so, and after a
large number of runs the inequality of (\ref{eq:Pinequality})
converges to an equality: \bq P_{\text{NA}}[D_i]_{\ket{\psi_{n+1}}}=
P_{\text{NA}}[D_i]_{\ket{\psi_{n}}}\quad\text{for
}n\to\infty.\label{eq:Pequalityjlarge} \eq

The equality in (\ref{eq:Pequalityjlarge}) is automatically satisfied
if $\ket{\psi_{n+1}}$ and $\ket{\psi_{n}}$ would differ only by a phase
factor, i.e.: \bq
\ket{\psi_{n+1}}=e^{i\alpha}\ket{\psi_{n}}\quad\text{for arbitrary
}e^{i\alpha}.\label{eq:samekets} \eq There is a class of internal
states that obey eq.~(\ref{eq:samekets}): the eigenstates of the
monodromy operator $\R^2$. If $\ket{\psi_{n}}$ would be an eigenstate of
$\R^2$ with eigenvalue $e^{i\l}$ then, after the $(n+1)$th run, the
internal state $\ket{\psi_{n+1}}$ would be, according to
(\ref{eq:stateevolution1}) if the outcome of the $n$th run had been
observation at detector $D_1$: \bq
\ket{\psi_{n+1}}=e^{i\alpha}\ket{\psi_{n}},\quad
e^{i\alpha}=\frac{1}{\sqrt{K}}\left(t_1r'_2e^{i\th_{\text{I}}}e^{i\l}+r_1t_2
e^{i\th_{\text{II}}}\right).\label{eq:lockedstate1} \eq If at the
$(n+1)$th run, particle $B$ was found at detector $D_2$, the new internal
state would have been, using (\ref{eq:stateevolution2}): \bq
\ket{\psi_{n+1}}=e^{i\alpha'}\ket{\psi_{n}},\quad
e^{i\alpha'}=\frac{1}{\sqrt{K'}}\left(t_1t'_2e^{i\th_{\text{I}}}e^{i\l} 
+r_1r_2
e^{i\th_{\text{II}}}\right).\label{eq:lockedstate2} \eq

The probability distributions $P_{\text{NA}}[D_i]_{\ket{\psi_n}}$
indeed turn out to converge to a fixed value that is associated with
an eigenvalue $e^{i\l}$ of the monodromy operator, when the number of
runs $n$ becomes large, i.e.: \bq \lim_{n\to\infty}
P_{\text{NA}}[D_1]_{\ket{\psi_n}}=|t_1r'_2|^2+|r_1t_2|^2+2Q \Re\left(
t_1r'_2r^*_1t^*_2e^{i\th}e^{i\l}\right).\label{eq:lockedP0} 
\eq In eq.~(\ref{eq:lockedP0}) we still used the notation that
includes the Mach-Zender apparatus' coefficients; let us write
(\ref{eq:lockedP0}) in the preferred device independent notation: \bq
\lim_{n\to\infty}
P_{\text{NA}}[D_i]_{\ket{\psi_n}}=P_{e^{i\l}}[D_i].\label{eq:lockedP1}
\eq

It is not only the probability distributions that converge, for the
internal state $\ket{\psi_n}$ also becomes fixed: as the eigenstate of
$\R^2$ that carries the eigenvalue $e^{i\l}$, in other words: \bq
\lim_{n\to\infty} \bra{\psi_n} \R^2 \ket{\psi_n}=e^{i\l} \eq So, we
conclude that {\em the internal state is locked into an eigenstate of
$\R^2$;} the only change subsequent runs of the experiment can
accomplish, is an uninteresting overall phase factor $e^{i\alpha}$, as
follows from (\ref{eq:lockedstate1}) and (\ref{eq:lockedstate2}).

But $\R^2$ has in general more than one eigenvalue. What can we say
about which eigenvalue $e^{i\l}$ will come out, i.e., to which
eigenvalue and eigenstate do $P_{NA}[D_i]_{\ket{\psi_n}}$ and
$\ket{\psi_n}$ will converge? This is where the initial internal state
$\ket{\psi}=\ket{\psi_0}$ comes in; if the expectation value of $\R^2$ and
$\ket{\psi}$ is given by: \bq \bra{\psi}\R^2\ket{\psi}=\sum_\l p_\l
e^{i\l}, \eq then the probability that the two particle system gets
locked in the eigenstate of $e^{i\l}$ is given by $p_\l$.

We will now consider the actual interference pattern
$I^{\text{o-o}}_{\text{NA}}[D_i]$ of the one-to-one (o-o) experiment
with non-abelian anyons and argue that it equals the fixed probability
distribution of (\ref{eq:lockedP1}). Let us assume that projection,
i.e., locking, is not achieved until the $m$th run. Then the outcomes
of the first $m$ runs were governed by other probabilities than of
eq.~(\ref{eq:lockedP1}) and the interference pattern, i.e., the
collection of outcomes at the time of the $m$th run, will not look
like (\ref{eq:lockedP1}). However, after the $m$th run, we can repeat
the experiment for an additional $n$ runs, where $n$ can be much
greater than $m$; all these subsequent outcomes will obey
(\ref{eq:lockedP1}) and, since $n\gg m$, completely dominate the
interference pattern, rendering the first $m$ contributions
negligible: \bq
\lim_{n\to\infty}I^{\text{o-o}}_{\text{NA}}[D_i]=P_{e^{i\l}}[D_i].
\label{eq:Ioneone}  
\eq

Not only does the monodromy operator $\R^2$ in general have multiple
distinct eigenvalues, these eigenvalues can be degenerate as
well. This degeneracy has no effect on the resulting interference
pattern, which remains of the form of (\ref{eq:Ioneone}). However, the
internal state does not become projected onto a particular eigenstate,
but rather onto the eigenspace of eigenvalue $e^{i\l}$: \bq
\lim_{n\to\infty}
\ket{\psi_n}=\frac{E_\l\ket{\psi}}{\sqrt{\bra{\psi}E_\l\ket{\psi}}},
\eq where the square root in the denominator is there to normalize the
state; $E_\l$ is the projection operator projecting onto the
eigenspace of $e^{i\l}$. The probability $p_\l$ can also be given in
terms of the projection operator $E_\l$: \bq
p_\l=\bra{\psi}E_\l\ket{\psi}.  \eq

We have showed that eigenstates of the monodromy operator are fixed
states, and said that all initial internal states eventually become
locked. The complete (nontrivial) proof of this convergence of
arbitrary initial states is rather lengthy, and we have relegated it
to the Appendix.  

Projection is achieved after a large but finite number of runs. How do
we know when we have reached such a projected, fixed, state? This, we
determine from the interference patterns: if we can identify the
interference pattern $I_{\text{NA}}^{\text{o-o}}[D_i]$ with extreme
certainty with one and only one $P_{e^{i\l}}[D_i]$, we know the
state is projected onto the eigenspace of $e^{i\l}$. The harder it is
to distinguish the $P_{e^{i\l}}[D_i]$ for different $e^{i\l}$, the
longer (more runs) it takes before projection will be realized.

Let us conclude the one-to-one experiment with an example, with the
same values (\ref{eq:coef_example}) for the coefficients of the
Mach-Zender interferometer, and a monodromy matrix $\R^2$ with
eigenvalues $\pm 1$, as we used before. In the following table 
we give the interference patterns $I^{\text{o-o}}_{\text{NA}}[D_i]$
and their probability to come out of the experiment for four different
initial internal states (which are characterized by $p_1$ and
$p_{=-1}$).  So, in this example, there are two possible 
observable interference patterns: either we observe nine out of ten
times particle $B$ at detector $D_1$ and one out of ten at detector
$D_2$ or the other way around (see the Table). Of the two situations
we may end up with, we can only calculate the probabilities, as it
should in this quantum mechanical setting.\\
\begin{quasitable}
\bc\parbox{9cm}{\narrowtext
\begin{tabular}{ccccc}
\tableline\tableline
$\Big.$$p_{1}$&$p_{-1}$&$I^{\text{o-o}}_{\text{NA}}[D_1]$&
$I^{\text{o-o}}_{\text{NA}}[D_2]$&Probability\\  
\tableline 
$\bigg.$$\case{1}{2}$ &$\case{1}{2}$ &$\case{9}{10}$
&$\case{1}{10}$ &$\case{1}{2}$\\ $\Big.$ & &$\case{1}{10}$
&$\case{9}{10}$ &$\case{1}{2}$\\

$\bigg.$$1$		&$0$		&$\case{9}{10}$
&$\case{1}{10}$	&$1$\\ 

$\bigg.$$0$		&$1$		&$\case{1}{10}$
&$\case{9}{10}$	&$1$\\ 

$\bigg.$$\case{3}{8}$	&$\case{5}{8}$	&$\case{9}{10}$
&$\case{1}{10}$	&$\case{3}{8}$\\ 
$\Big.$		&		&$\case{1}{10}$	&$\case{9}{10}$
&$\case{5}{8}$\\ 
\tableline\tableline
\end{tabular}\widetext}\ec
\end{quasitable}

\subsection{The many-to-one experiment: probe one particle with many others}

In the many-to-one experiment, the experiment is repeated with the
same particle $A$ but with a new particle $B$ each time. This will
lead to a different result, for now the system will get locked into an
eigenstate of a different operator called $U$, which will be defined
below. 

\begin{figure}
\bc\parbox{9cm}{\narrowtext
\includegraphics{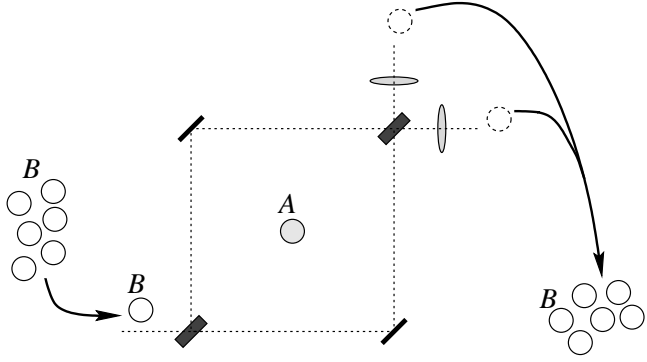}\\
\caption{The {\em many-to-one} experiment: the same particle $A$ remains
continuously at the center of the apparatus during the experiment, but
particle $B$ gets replaced by a new one after detection. All $B$
particles initially carry the same internal state. The multi-particle
internal state changes with each observation of the current particle
$B$ at detector $D_1$ or $D_2$ and may become highly entangled.}
\label{fig:MZ_mo}
\widetext}\ec
\end{figure}

The setup of the Mach-Zender apparatus in this case no longer
requires the extension with mirrors as needed for the one-to-one
experiment, but it does require a large supply, i.e., a `bag' with many
identical particles $B$ in it, on the left of the apparatus,
see Fig.~\ref{fig:MZ_mo}.
We will be using the notation/convention for braids of these multiple
particles as we introduced in section \ref{sec:intro} (i.e. writing
particles left in the experiment left in the tensor product, $\R_i$
interchanging particles $i$ and $(i+1)$ anticlockwise, numbering form
right to left beginning with one). 
The total system's initial internal state $\ket{\psi}$ is an element
of the tensor product of the internal spaces $V^B$ of the $B$
particles and $V^A$ of the $A$ particle: \bq \ket{\psi}\in 
V^B\otimes \cdots \otimes V^B\otimes V^B\otimes V^B\otimes V^A. \eq
Once a $B$ particle has been used, it has become uninteresting but we
will not discard 
it completely, merely collect such $B$ particles somewhere on the
right of the 
apparatus.

For reasons of simplicity and to give a full solution, we will subject
the internal state $\ket{\psi}$ to the following conditions: the
initial internal state $\ket{\psi}$ should factorize completely, i.e.,
the initial internal state should be \emph{unentangled}, and
furthermore, the $B$ 
particles should have trivial braiding: \bq
\R_i\ket{\psi}=\RI_i\ket{\psi}=\sigma_i\ket{\psi}=\ket{\psi}
\quad\quad\forall\,  
i\ge2, \eq where $\sigma_i$ is the operator that only swaps particle
$i$ and $(i+1)$ without affecting their internal state. 
Applying both conditions to the internal
state $\ket{\psi}$, implies that $\ket{\psi}$ can be written in the
following form: \bq \ket{\psi}=y^{i_n}\ldots y^{i_2}y^{i_1}
x^{\alpha}\ket{e_{i_n}}\ldots\ket{e_{i_2}}\ket{e_{i_1}}\ket{f_\alpha},
\eq where $\{\ket{e_{i_j}}\}$is a basis for the internal space $V^B$
of the $j$th particle $B$, and $\{\ket{f_\alpha}\}$ a basis for $V^A$.

Let us start the first run of the thought experiment: we insert the
first particle $B$ in the Mach-Zender interferometer and observe it
again at either detector $D_1$ or $D_2$. The probabilities
$P_{\text{NA}}[D_i]$ for detector $D_i$ are of course given by
(\ref{eq:Pna1b}), (\ref{eq:Pna2b}), and depend on the expectation
value $\bra{\psi_0}\R^2_1\ket{\psi_0}$ of the monodromy operator
$\R^2_1$ which acts on particle $A$ and the first particle $B$, with
respect to the initial internal state
$\ket{\psi_0}\equiv\ket{\psi}$. Let us 
assume we observe the first incident particle at detector $D_1$, then
the new, changed, internal state, which we indicate by $\ket{\psi_1}$,
is given by: \bq 
\ket{\psi_1}=\frac{1}{\sqrt{K}}\left(t_1r'_2e^{i\th_{\text{I}}}\R_1+r_1t_2
e^{i\th_{\text{II}}}\RI_1\right)\ket{\psi_0}.\label{eq:MOstateevolution}
\eq Because the first particle $B$ is now to the right of particle
$A$, the new internal state $\ket{\psi_1}$ is strictly speaking an
element of a different vector space than that of $\ket{\psi_0}$: \bq
\ket{\psi_1}\in V^B\otimes \cdots \otimes V^B\otimes V^B\otimes
V^A\otimes V^B.  \eq Furthermore, the first particle $B$ and particle
$A$ are now entangled, so the internal state $\ket{\psi_1}$ cannot be
completely factorized anymore: \bq \ket{\psi_1}=y^{i_n}\ldots
y^{i_2}z^{\alpha
i_1}\ket{e_{i_n}}\ldots\ket{e_{i_2}}\ket{f_\alpha}\ket{e_{i_1}}, \eq
where the components $z^{\alpha i_1}$ are determined by the
coefficients of (\ref{eq:MOstateevolution}) and the components
$x^\alpha$ and $y^i$. The second up to the $n$th $B$ particle, still
have trivial braiding, which we can generalize to the $j$th run of the
experiment 
 by the statement: \bq
\R_i\ket{\psi_j}=\ket{\psi_j},\quad\quad
i>j+1.\label{eq:trivialbraiding2} \eq So far for the first run.

For the second run, the next particle $B$ is injected in the
Mach-Zender apparatus, and the probabilities to observe it at detector
$D_i$ are now governed by the expectation value
$\bra{\psi_1}\R_2^2\ket{\psi_1}$ of the current internal state
$\ket{\psi_1}$ and the monodromy operator $\R_2^2$ that operates on
particle $A$ and the second particle $B$. This expectation value will
in general be different from the expectation value of the first run:
$\bra{\psi_1}\R_2^2\ket{\psi_1}\ne\bra{\psi_0}\R_1^2\ket{\psi_0}$, and
thus the probabilities for the first and second run will be
different. But, just as in the case of the one-to-one experiment, the
probabilities and thus the expectation values will converge to a
final, fixed, value: \bq
\bra{\psi_{n+1}}\R_{n+2}^2\ket{\psi_{n+1}}=\bra{\psi_{n}}\R_{n+1}^2
\ket{\psi_{n}},\quad\quad  
n\to\infty.\label{eq:MOexpectationlimit} \eq We would like to know
which states obey this equality (\ref{eq:MOexpectationlimit}) and what
the expectation value will be. These will not be eigenstates and
eigenvalues of a monodromy operator, as for the one-to-one experiment,
but something related.

Let us now start solving equation (\ref{eq:MOexpectationlimit}) if we
assume that the $(n+1)$th particle $B$ has been observed at detector
$D_1$, and use that we know $\ket{\psi_{n+1}}$ in terms of
$\ket{\psi_{n}}$, as in (\ref{eq:MOstateevolution}): \widetext \bqa
\bra{\psi_{n+1}}\R_{n+2}^2\ket{\psi_{n+1}}&=&\frac{1}{K}\biggl(\,
|t_1r'_2|^2\bra{\psi_{n}}\RI_{n+1}\R^2_{n+2}\R_{n+1}\ket{\psi_{n}}
+t_1r'_2r_1^*t_2^* 
e^{i(\th_{\text{I}}-\th_{\text{II}})}\bra{\psi_{n}}\R_{n+1}\R^2_{n+2}
\R_{n+1}\ket{\psi_{n}}\nn\\  
&&\quad\quad+|r_1t_2|^2\bra{\psi_{n}}\R_{n+1}\R^2_{n+2}\RI_{n+1}
\ket{\psi_{n}} 
+t_1^*r'^*_2r_1t_2
e^{-i(\th_{\text{I}}-\th_{\text{II}})}\bra{\psi_{n}}
\RI_{n+1}\R^2_{n+2}\RI_{n+1}\ket{\psi_{n}} 
\biggr)\label{eq:solveMO1}\\ &=&\bra{\psi_{n}}\R_{n+1}^2\ket{\psi_{n}},\nn
\eqa where the normalization constant $K$ is given by: \bq
K=|t_1r'_2|^2+|r_1t_2|^2+2 \Re\left(t_1r'_2r_1^*t_2^*
e^{i(\th_{\text{I}}-\th_{\text{II}})}\bra{\psi_{n}}\R_{n+1}^2\ket{\psi_{n}}
\right)\label{eq:solveMO2}.  \eq 
These two expressions,
(\ref{eq:solveMO1}), (\ref{eq:solveMO2}), may look horrid (and we can
write down another pair for the case that outcome of the $(n+1)$th run is
observation at detector $D_2$), but they are not that bad if we apply
the braid relations to them, i.e., use the fact that the $\R_n$ satisfy
the Yang-Baxter relations. Using
(\ref{eq:trivialbraiding2}) as well, two of the four
expectation values in (\ref{eq:solveMO1}) simplify to: \bqa
\bra{\psi_{n}}\RI_{n+1}\R^2_{n+2}\R_{n+1}\ket{\psi_{n}}&=&\bra{\psi_{n}} 
\R_{n+2}\R^2_{n+1}\RI_{n+2}\ket{\psi_{n}}\nn\\
&=&\bra{\psi_{n}}\R^2_{n+1}\ket{\psi_{n}}\nn\\ &\equiv&\kappa, \eqa \bqa
\bra{\psi_{n}}\R_{n+1}\R^2_{n+2}\RI_{n+1}\ket{\psi_{n}}&=&\bra{\psi_{n}} 
\RI_{n+2}\R^2_{n+1}\R_{n+2}\ket{\psi_{n}}\nn\\
&=&\bra{\psi_{n}}\R^2_{n+1}\ket{\psi_{n}}\nn\\ &=&\kappa, \eqa where we
introduced $\kappa$ to indicate the expectation value
$\bra{\psi_{n}}\R^2_{n+1}\ket{\psi_{n}}$. Equation (\ref{eq:solveMO1}) will
be solved if $\ket{\psi_{n}}$ is such that: \bq
\bra{\psi_{n}}\R_{n+1}\R^2_{n+2}\R_{n+1}\ket{\psi_{n}}=\kappa^2
\label{eq:solveMO3}, 
\eq \bq \bra{\psi_{n}}\RI_{n+1}\R^2_{n+2}\RI_{n+1}\ket{\psi_{n}}=
\kappa\kappa^*. 
\eq Let us rewrite these expectation values too, by applying then
Yang-Baxter equation and eq.~(\ref{eq:trivialbraiding2}) once more:
\bq
\bra{\psi_{n}}\R_{n+1}\R^2_{n+2}\R_{n+1}\ket{\psi_{n}}=\bra{\psi_{n}}
\R_{n+1}^2\sigma_{n+2}\R^2_{n+1}\sigma_{n+2}\ket{\psi_{n}}
\label{eq:MOexpectationvalue1},  
\eq \bq
\bra{\psi_{n}}\RI_{n+1}\R^2_{n+2}\RI_{n+1}\ket{\psi_{n}}=\bra{\psi_{n}} 
\R_{n+1}^{-2}\sigma_{n+2}\R^2_{n+1}\sigma_{n+2}\ket{\psi_{n}}
\label{eq:MOexpectationvalue2}. 
\eq

Although $\R_{n+1}^2$ and $\sigma_{n+2}\R^2_{n+1}\sigma_{n+2}$ are distinct
operators, their matrix elements $(R^2)^{i \alpha}_{j\beta}$ are equal:
\bq
(R^2)^{i \alpha}_{j\beta}=\bra{e^i}\bra{f^\alpha}\R^2\ket{e_j}\ket{f_\beta}.
\eq

We will decompose the right hand side of
(\ref{eq:MOexpectationvalue1}) in a similar way in components relative
to the bases $\{\ket{e_i}\}$, $\{\ket{f_\alpha}\}$ of the vector
spaces $V^B$ and $V^A$. The internal state $\ket{\psi_{n}}$ decomposes
as: \bq \ket{\psi_{n}}=y^i y^j z^{\alpha i_{n}\ldots
i_1}\ket{e_i}\ket{e_j}\ket{f_\alpha}\ket{e_{i_{n}}}\ldots\ket{e_{i_1}}.
\eq Next, we let the combined operators $\R_{n+1}^2$ and
$\sigma_{n+2}\R^2_{n+1}\sigma_{n+2}$ act on a basis state (ignoring the
$\ket{e_{i_{n}}}\ldots\ket{e_{i_1}}$ part) to give: \bq
\R_{n+1}^2\sigma_{n+2}\R^2_{n+1}\sigma_{n+2}\ket{e_i}\ket{e_j}\ket{f_\alpha}
=(R^2)^{k\beta}_{i\alpha}(R^2)^{l\gamma}_{j\beta}\ket{e_k}\ket{e_l}
\ket{f_\gamma}. 
\eq If we also introduce the density matrices $\rho_A$ and $\rho_B$ of
particles $A$ and $B$: \bq (\rho_A)^\alpha_\beta=z^{\alpha
i_{n}\ldots i_1}z_{\beta i_{n}\ldots
i_1},\quad\quad(\rho_B)^i_j=y^i y_j, \eq then the expectation value of
(\ref{eq:MOexpectationvalue1}) can be written as follows: \bqa
\bra{\psi_{n}}\R_{n+1}^2&&\sigma_{n+2}\R^2_{n+1}\sigma_{n+2}\ket{\psi_{n}}
=\\
&&=(R^2)^{k\beta}_{i\alpha} (R^2)^{l\gamma}_{j\beta} (\rho_B)^i_k
(\rho_B)^j_l (\rho_A)^\alpha_\gamma\nn\\
&&=[(\rho_B)^j_l(R^2)^{l\gamma}_{j\beta}][
(\rho_B)^i_k(R^2)^{k\beta}_{i\alpha}](\rho_A)^\alpha_\gamma\nn\\ 
&&=\U^\gamma_\beta\U^\beta_\alpha (\rho_A)^\alpha_\gamma\nn\\
&&=\Tr(UU\rho_A).\label{eq:Uintro1} \eqa Here, in the last two steps
of eq.~(\ref{eq:Uintro1}), we introduced -- as announced -- an
operator $U$ that operates 
on $V^A$ and is defined as the partial trace over the $B$ space of the
monodromy operator $\R^2$ and the density matrix of particle $B$: \bq
\U^\alpha_\beta=(\rho_B)^j_i(R^2)^{i\alpha}_{j\beta},\quad\quad
\U_A=\Tr_B(\rho_B \R^2_{BA}).  \eq

Let us now reconsider our original problem in terms of the $\U$
and $\kappa$. We can express $\kappa$ as the trace over $\U$ and the
density matrix of particle $A$: \bq \kappa=\Tr(\U\rho_A).  \eq The
condition to satisfy equality (\ref{eq:solveMO1}), as given by
(\ref{eq:solveMO3}), becomes in terms of $\U$: \bq
\Tr(\U^2\rho_A)=\kappa^2.  \eq Since the trace of an operator and a
density matrix can be regarded as the expectation value of that
operator, we have that: \bq \langle\U^2\rangle=\langle\U\rangle^2, \eq
which can only be true if $\kappa$ is an eigenvalue of $\U$ and
$\rho_A$ lies in the eigenspace of the eigenvalue $\kappa$.

Just as in the one-to-one experiment, the probability distributions
for the incident particles converge to a fixed distribution associated
with the eigenvalue of an operator, and the final distribution
determines the interference pattern. Also, the system becomes
projected onto the eigenspace of the observed eigenvalue. However, the
eigenvalues and eigenstates are no longer given by the monodromy
operator $\R^2$ as in the one-to-one experiment; in the many-to-one
(m-o) experiment we observe an eigenvalue $\kappa$ of the operator
$\U$ and have projection onto the eigenspace of that eigenvalue. If we
denote the probability distributions by
$P^{\text{m-o}}_{\text{NA}}[D_i]$, we see that for large $n$ they
become equal to the interference patterns
$I^{\text{m-o}}_{\text{NA}}[D_i]$ : \bqa
\lim_{n\to\infty}P^{\text{m-o}}_{\text{NA}}[D_1]_{\ket{\psi_{n}}}&=&
\lim_{n\to\infty}I^{\text{m-o}}_{\text{NA}}[D_1]\nn\\ 
&=&|t_1r'_2|^2+|r_1t_2|^2+2Q
\Re\left(t_1r'_2r^*_1t^*_2e^{i\th}\kappa\right)\nn\\
&\equiv&P_{\kappa}[D_1], \eqa where we used $P_{\kappa}[D_1]$ to
indicate the locked probability distribution that is associated with
the eigenvalue $\kappa$; of course there are similar expressions for
$D_2$. As far as the projection of the internal state is concerned,
there is strictly speaking only a projection in a subsystem of the
total system's internal state, namely the subsystem describing the
internal state of particle $A$, because $\U$ acts on $V^A$ only. It is
the density matrix of particle 
$A$ that becomes projected onto the associated eigenspace of $\kappa$:
\bq \rho_A\to\frac{E_\kappa \rho_A E_\kappa}{\Tr(E_\kappa
\rho_A)}\quad\quad\text{for }n\to\infty, \eq where $E_\kappa$ is the
projection operator of the eigenspace of $\kappa$. Just as the
monodromy operator $\R^2$ can have multiple (degenerate) eigenvalues,
so can $\U$. Each eigenvalue $\kappa$ has a probability $p_\kappa$ to
be observed in the end; $p_\kappa$ is determined by the initial
density matrix $\rho_A$: \bq p_\kappa=\Tr(E_\kappa
\rho_A),\quad\quad(\rho_A)^\alpha_\beta=x^\alpha x_\beta.  \eq

Above, we have showed that there exist states for which the
probability distributions are locked, and that there is a chance
$p_\kappa$ to end up in each locked state. The proof of it is almost
the same as that of the convergence in the one-to-one experiment and
can be found in the Appendix. 

To solve the many-to-one experiment we introduced two conditions: the
initial internal state should factorize completely and the $B$
particles should have trivial braiding. We believe these conditions
are too restrictive: the many-to-one experiment should be possible for
arbitrary entangled initial states, subjected only to the condition
that the $B$ particles have abelian braiding: \bq
\R_j\ket{\psi_k}=e^{is_B}\ket{\psi_k}\quad\quad
j>k,\label{eq:generalMO1} \eq where the $s_B$ in the phase factor
$e^{is_B}$ is also known as the spin of the $B$ particles. In this
generalized many-to-one experiment, the interference patterns also
depend explicitly on $\kappa$: \bq
\lim_{n\to\infty}I^{\text{m-o}}_{\text{NA}}[D_i]=P_{\kappa}[D_i],
\label{eq:generalMO2}  
\eq but $\kappa$ can no longer be regarded as the eigenvalue of the
operator $\U$; the definition of $\U$ becomes ambiguous, because $\U$
is the trace of $\R^2$ and the density matrix $\rho_B$ of the $B$
particles, and in this case $\rho_B$ is no longer constant. We have
not yet succeeded in proving this for the generalized many-to-one
experiment, so the claim of (\ref{eq:generalMO2}) with the condition
of (\ref{eq:generalMO1}) remains for the moment a \emph{conjecture}.%
\footnote{To prove eq.~(\ref{eq:generalMO2}) with
(\ref{eq:generalMO1}) certainly requires some additional properties of
non-abelian anyons and physical restrictions on the allowed internal
states, especially concerning the creation and fusion of non-abelian
anyons, for which the description of non-abelian anyons through a
quantum group may be instrumental. This is beyond the scope of the
present paper.}

Let us now give an example to illustrate the many-to-one experiment,
which is borrowed from an example in \cite{BJOscriptie}. We will take
 the internal space $V^A$ of the $A$ particle to be three-dimensional
and to have a basis $\{\ket{1},\ket{2},\ket{3}\}$. The internal space
$V^B$ will be two-dimensional and it will have a basis
$\ket{+},\ket{-}$. The monodromy operator $\R^2$ is thus
six-dimensional. Relative to the chosen bases, the monodromy operator
we will use is, in a basis
$\{\ket{+}\ket{1},\ket{-}\ket{1},\ket{+}\ket{2},\ket{-}\ket{2},
\ket{+}\ket{3},\ket{-}\ket{3}\}$, 
given by:\\ \bq \R^2= \left(\begin{array}{ccc} 
\begin{array}{cc}
-\frac{1}{2}&\frac{1}{2}\sqrt{3}\\
\frac{1}{2}\sqrt{3}&\frac{1}{2}
\end{array}&0&0\\
0&
\begin{array}{cc}
1&0\\0&-1
\end{array}
&0\\
0&0&\begin{array}{cc}
-\frac{1}{2}&-\frac{1}{2}\sqrt{3}\\
-\frac{1}{2}\sqrt{3}&\frac{1}{2}
\end{array}
\end{array}\right)
.\label{eq:explicitR2}
\eq \\
This specific $\R^2$ has two eigenvalues: $+1,-1$, which are both
three-fold degenerate. Of course, it is a unitary matrix.  We choose
the incident $B$ particles to be in the state $\ket{+}$, which means
that their density matrix in a basis $\{\ket{+},\ket{-}\}$ is: \bq
\rho_B=\left(\begin{array}{cc}1&0\\0&0\end{array}\right)
.
\eq The matrix of the operator $\U$ can be simply obtained by
performing the partial trace of $\R^2$ and $\rho_B$, in a basis
$\{\ket{1},\ket{2},\ket{3}\}$: \bq 
\U=\Tr_B(\R^2\rho_B)=\left(\begin{array}{ccc}-\frac{1}{2}&0&0\\0&
1&0\\0&0&-\frac{1}{2}\end{array}\right)  
.
\eq So, there are two different eigenvalues $\kappa$: $+1$ and
$-\frac{1}{2}$. If we now plug in the values for the Mach-Zender
apparatus' coefficients, (\ref{eq:coef_example}), in the interference
patterns $I^{\text{m-o}}_{\text{NA}}[D_i]$ associated with each
eigenvalue, these interference patterns take the form given
in the Table below.
One outcome of this example of the many-to-one experiment is equal to
the outcome of the ordinary interference experiment and the one-to-one
experiment: nine out of ten times the incident particle is found at
detector $D_1$. The other possible outcome is rather different: three
out of ten times is is detected by detector $D_1$ and with a 70\%
chance the $B$ particle is observed at detector $D_2$.\\

\begin{quasitable}\bc\parbox{9cm}{\narrowtext
\begin{tabular}{ccc}
\tableline\tableline
$\big.$&\multicolumn{2}{c}{Interference
patterns}\\ Eigenvalue
$\kappa$&$I^{\text{m-o}}_{\text{NA}}[D_1]$&$I^{\text{m-o}}_{\text{NA}} 
[D_2]$\\
\tableline $\Big.$$+1$&$\frac{9}{10}$&$\frac{1}{10}$\\
$\Big.$$-\frac{1}{2}$&$\frac{3}{10}$&$\frac{7}{10}$\\
\tableline\tableline
\end{tabular}\widetext}\ec
\end{quasitable}

The above example also demonstrates that the eigenvalues of the
operator $\U$ need not have an absolute value of one. This would have
been the case if $\U$ were a unitary operator, but $\U$ is merely a
{\em normal} operator, i.e., it commutes with its adjoint: \bq
\U\U^\dag=\U^\dag\U, \eq which follows from the unitarity of $\R^2$
and the hermicity of density matrices.  Because of the definition of
$\U$ in terms of the monodromy operator $\R^2$, one can easily deduce
that $\kappa$ can be written as a linear combination of eigenvalues
$e^{i\l}$ of $\R^2$ and that its absolute value cannot
exceed one: \bq \kappa=\sum_\l p_\l e^{i\l},\quad\quad\sum_\l
p_\l=1,\quad\Rightarrow\quad |\kappa|\le1.  \eq

\subsection{Summary}
Let us summarize the results of the three different thought
experiments we have described in this paper: the many-to-many, one-to-one,
and many-to-one experiments. Table \ref{table:summary}\ gives a
compact survey of the results.

\begin{table}[!t]
\caption{Summary of the results for the three different thought
interference experiments with non-abelian anyons we discussed. Notice
that the results are independent of the exact setup of the Mach-Zender
interferometer and may be applied to other setups as well. The
one-to-one and many-to-one experiment lock the system into an
eigenpattern or eigenvalue associated with an operator (either the
monodromy operator $\R^2$, or $\U$: the partial trace over $\R^2$ and
the density matrix of particle $B$) acting on the internal state, and
project the internal state onto the eigenspace. The many-to-many
experiment merely determines the expectation value of the monodromy
operator $\R^2$, and ignores any change in the internal state due to
the measurements. Although interference patterns cannot be identified
with probability distributions, there is still some conservation of
probability: the average of the outcomes of all three different
schemes is the same. For ordinary interference experiments or
Aharonov-Bohm experiments there is no need to distinguish between the
many-to-many, the one-to-one and the many-to-one schemes, as all
schemes will yield the same results in these cases, the difference
becomes important for non-abelian anyons only.}\label{table:summary}
\begin{tabular}{lccc}
&\multicolumn{3}{c}{type of interference
experiment}\\ &many-to-many&one-to-one&many-to-one%
\tablenote{Recall
that to solve the many-to-one experiment we posed some restrictions on
the internal state of the $B$ particles; the operator $\U$ can only be
defined if the initial internal state is completely unentangled.} \\
\tableline \\ The measured expectation/eigen value,
&$\bra{\psi}\R^2\ket{\psi}$&$e^{i\l}$&$\kappa$\\
$\bra{\psi_{n}}\R^2_{n+1}\ket{\psi_{n}}$ for 
$n\to\infty$\\ \\ The diagonalised operator &not
relevant&$\R^2$&$\U=\Tr_B(\R^2\rho_B)$\\ \\ The 
observed interference  pattern, &$\sum_\l p_\l
P_{e^{i\l}}[D_i]$&$P_{e^{i\l}}[D_i]$&$P_\kappa[D_i]$\\
$I_{\text{NA}}[D_i]=\lim_{n\to\infty}
P_{\text{NA}}[D_i]_{\bra{\psi_{n}}\R^2_{n+1}\ket{\psi_{n}}}$\\ \\ Change of
internal state&not relevant&$\ket{\psi}\to\frac{\displaystyle E_\l
\ket{\psi}}{\displaystyle\sqrt{\bra{\psi}E_\l\ket{\psi}}}$&
$\rho_A\to\frac{\displaystyle 
E_\kappa\rho_A E_\kappa}{\displaystyle\Tr_A(E_\kappa\rho_A)}$\\ \\
Probability of
eigenvalue/pattern&$1$&$p_\l=\bra{\psi}E_\l\ket{\psi}$&
$p_\kappa=\Tr_A(E_\kappa 
\rho_A)$\\ \\ Average of all outcomes &
$\bra{\psi}\R^2\ket{\psi}$&$\sum_\l p_\l
e^{i\l}=\bra{\psi}\R^2\ket{\psi}$&$\sum_\kappa p_\kappa \kappa
=\bra{\psi}\R^2\ket{\psi}$\\ \\ \tableline \\ Aharonov-Bohm pattern
$I_{\text{AB}}[D_i]$
&$P_{e^{i\l}}[D_i]$&$P_{e^{i\l}}[D_i]$&$P_{e^{i\l}}[D_i]$\\ 
with $\R^2=e^{i\l}\openone$\\ \\
The plain interference pattern $I[D_i]$
&$P_{1}[D_i]$&$P_{1}[D_i]$&$P_{1}[D_i]$\\ with $\R^2=\openone$\\ \\
\end{tabular}
\end{table}

The topological interactions of non-abelian anyons are generally
agreed upon to be observable in interference experiments. We have
demonstrated that such interference experiments are generalizations of
interference experiments of the Aharonov-Bohm type. However, whereas
in the Aharonov-Bohm versions, the probability distributions are
identified with the interference patterns, this cannot be readily
carried over to experiments involving non-abelian anyons. There it
is crucial to explicitly describe the way in which each run of
an experiment is repeated.

In the many-to-many experiment, each new run is conducted with fresh
particles $A$ and $B$ carrying the initial internal state
$\ket{\psi}$. In this case, by definition, the interference pattern
reproduces the probability distribution, and is characterized by the
expectation value $\bra{\psi}\R^2\ket{\psi}$ of the monodromy operator
$\R^2$.

The scheme of the one-to-one experiment uses the same two particles
$A$ and $B$ over and over again. The resulting interference pattern
now does not depend on $\bra{\psi}\R^2\ket{\psi}$, but rather on an
eigenvalue $e^{i\l}$ of the unitary monodromy operator $\R^2$. Because
there is a clear analogy between the eigenvalues $e^{i\l}$ and the
associated interference patterns $P_{e^{i\l}}[D_i]$ we could  use the
term \emph{eigenpattern} to indicate the eigenvalue-pattern
$P_{e^{i\l}}[D_i]$. So, in the one-to-one experiment we observe an
eigenpattern, or in other words we measure an eigenvalue. The
measurement of the eigenvalue projects the two-particle internal state
onto the eigenspace of $e^{i\l}$. The probability to observe a
specific eigenpattern is determined by $p_\l$: the fraction of the
internal state $\ket{\psi}$ that carries eigenvalue $e^{i\l}$.

The many-to-one experiment describes the way one intuitively (and
conventionally) envisages the repetition of an interference experiment
with non-abelian anyons: fix particle $A$ inside the interferometer
and direct many identically prepared particles $B$ at it. However, the
many-to-one turns out mathematically to be the hardest of all three
schemes to analyze, as we need to deal with the total system's internal
state each run. If we restrict ourselves to initially unentangled
internal states, we can identify the resulting eigenpattern
$P_\kappa[D_i]$ with the eigenvalue $\kappa$ of an operator $\U$ that
acts on the internal space of particle $A$; the operator $\U$
explicitly depends on the internal state of particle $B$, as $\U$ is
the partial trace over $\R^2$ and $\rho_B$. The possible eigenvalues
$\kappa$ of $\U$ thus also depend on $\rho_B$. The projection of the
system onto the eigenspace of $\kappa$ effectively only affects the
$A$ particle: it is its density matrix $\rho_A$ that gets projected.

Interference experiments, including thought experiments, obviously
need a device that can create interference; the topological
interaction of the non-abelian anyons does not cause interference, it
only alters existing interference. We chose to use the Mach-Zender
interferometer and explicitly used its determining coefficients $t_1$,
$r'_2$, $Q$ etc., in many equations and our examples. In the end
however, we switched to a notation that is independent of the exact
setup, and only depends on eigenvalues and eigenpatterns. The results
in Table~\ref{table:summary} are thus not restricted to experiments
with the Mach-Zender interferometer, but apply to interference
experiments in general, for instance double slit experiments.

Although the three schemes yield different results, the probability
distribution for the first incident particle is the same for each
scheme, and we expect to see some conservation of this initial
probability. This is achieved by taking the \emph{average} of all
outcomes. When averaged, the many-to-many, one-to-one, and many-to-one
experiments all return the same result which reflects the equal
probability for the first run of either experiment.

Being a generalization, the non-abelian anyon experiments should of
course cover the Aharonov-Bohm effect and the ordinary interference
experiments as well. If we let the monodromy operator act on a trivial
internal space, i.e., $\R^2=e^{i\l}\openone$ or just $\R^2=e^{i\l}$,
we retrieve the Aharonov-Bohm effect: the interference pattern for all
three schemes will be the same $P_{e^{i\l}}[D_i]$, and the question of
projection becomes irrelevant on the trivial internal space. If we let
$\R^2$ act as unity, we recover the ordinary interference experiment,
and the three different ways to repeat the experiment obviously yield
the same interference patterns. The difference between interference
patterns and probability distributions becomes non-trivial \emph{only}
in the case of non-abelian anyons.

\appendix

\section*{Proof of locking}
Here in the Appendix, we will give the proof that in the one-to-one
and many-to-one experiments, all initial internal states converge to a
locked state. We will begin from the point of view of the one-to-one
experiment, and we will see later on that the proof for the
many-to-one experiment is essentially the same. We will repeat some of
the more important equations from sections \ref{sec:MZsetup} and
\ref{sec:experiments}, and we will state what needs to be proven from
the physical point of view. From there on, the proof is purely
mathematical, and we will change notation as to clearly discriminate
physics from mathematics. The pivot of proof is to construct a new
probability distribution function, which has a clear behavior as a
function on $n$, and shows that $\ket{\psi_n}$ becomes projected for
large $n$ with the correct probability. 

So, focusing on the one-to-one experiment now, we will first determine
an expression for $\bra{\psi_n}E_\mu\ket{\psi_n}$ as a function of the
outcomes $D_{i_j}$ ($j\in\{1,\ldots,n\}$, $i_j\in\{1,2\}$) of the
performed runs of the experiment. If the two-particle internal state
after the $j$th run is $\ket{\psi_{j}}$, then based on the outcome of
the $(j+1)$th run, $D_{i_{j+1}}$, the new internal state
$\ket{\psi_{j+1}}$ becomes,
eqs. (\ref{eq:stateevolution1}),(\ref{eq:stateevolution2}): 
\bq\label{eq:app:psin1}
\ket{\psi_{j+1}}=\frac{1}{\sqrt{K'}}(t_1r'_2e^{i\th_{\text{I}}}\R^2
+r_1t_2e^{i\th_{\text{II}}}\openone)\ket{\psi_{j}},  
\eq
when the outcome is observation at detector $D_1$; if the incident
particle emerges at detector $D_2$, the new state is given by: 
\bq\label{eq:app:psin2}
\ket{\psi_{j+1}}=\frac{1}{\sqrt{K''}}(t_1t'_2e^{i\th_{\text{I}}}
\R^2+r_1r_2e^{i\th_{\text{II}}}\openone)\ket{\psi_{j}}. 
\eq
The dependence on the outcome $D_{i_{j+1}}$ is not explicit yet in the
above expressions. However, we can rewrite the parts
$(\ldots\R^2+\ldots\openone)$ in terms of $E_\l$ and $e^{i\l}$, if we
use that (eqs. (\ref{eq:El1}),(\ref{eq:El2})): 
\[
\R^2=\sum_\l e^{i\l} E_\l,\quad\quad\openone=\sum_\l E_\l,
\]
for then:
\bqa
t_1r'_2e^{i\th_{\text{I}}}\R^2+r_1t_2e^{i\th_{\text{II}}}\openone&=&
\sum_\l(t_1r'_2e^{i\th_{\text{I}}}e^{i\l}
+r_1t_2e^{i\th_{\text{II}}})E_\l\nn\\  
&\equiv&\sum_\l f_\l[D_1] E_\l\nn.
\eqa
Here, we introduced $f_\l[D_i]$ for the sake of abbreviation, but also
the dependence on both the eigenvalues $e^{i\l}$ and the outcomes
$D_{i}$ becomes more clear. 
If we take the absolute square of $f_\l[D_i]$ (and implicitly
integrate over $q$) we recognize the Aharonov-Bohm probability
distribution $P_{e^{i\l}}[D_i]$: 
\[
\big| t_1r'_2e^{i\th_{\text{I}}}e^{i\l}+r_1t_2e^{i\th_{\text{II}}}
\big|^2=\big|f_\l[D_1]\big|^2=P_{e^{i\l}}[D_1]. 
\]
If we write $\ket{\psi_{n}}$ in terms of $\ket{\psi_0}$, according to
eqs.~(\ref{eq:app:psin1}), (\ref{eq:app:psin2}) and  by substituting
$f_\l[D_i]$,  we obtain a product of $n$ similar factors: 
\[
\ket{\psi_{n}}=\frac{1}{\sqrt{K}}\!\left(\sum_{\l_n}f_{\l_n}[D_{i_n}]
E_{\l_n}\!\!\right)\!\ldots\!\left(\sum_{\l_1}f_{\l_1}[D_{i_1}]
E_{\l_1}\!\!\right)\!\ket{\psi_0}. 
\]
Most of the projectors $E_{\l_j}$ cancel each other, if we apply to
them the projector algebra: 
\[
E_\l E_\mu=\delta_{\l\mu} E_\l.
\]

Now, we are ready to express the expectation value
$\bra{\psi_{n}}E_\mu\ket{\psi_{n}}$ with explicit dependence on the
outcomes $\{D_{i_j}\}$ of the $n$ runs, in the following way: 
\bq\label{eq:En_expec1}
\bra{\psi_{n}}E_\mu\ket{\psi_{n}}=\frac{1}{K}P_{e^{i\mu}}[D_{i_1}]\ldots
P_{e^{i\mu}}[D_{i_n}]\,\bra{\psi_0}E_\mu\ket{\psi_0}. 
\eq
The normalization constant $K$ is simply given by:
\bqa
K&=&\sum_\l P_{e^{i\l}}[D_{i_1}]\ldots
P_{e^{i\l}}[D_{i_n}]\,\bra{\psi_0}E_\l\ket{\psi_0}\\ 
&\equiv&p_n(D_{i_1},D_{i_2},\ldots,D_{i_n}),
\eqa
but this $K$ turns out to have a physical meaning too: it is equal to
the probability $p_n(D_{i_1},D_{i_2},\ldots,D_{i_n})$ to observe the
particular sequence of $n$ outcomes at the detectors $D_{i_1}$,
$D_{i_2}$, \ldots, $D_{i_n}$. With this probability in mind, proving
that the state locks means we need to show that for large values of
$n$ the following is true for a certain eigenvalue $e^{i\mu}$: 
\bqa
\bra{\psi_{n}}E_\mu\ket{\psi_{n}}&\to& 1,\nn\\
\bra{\psi_{n}}E_\l\ket{\psi_{n}}&\to& 0,\quad \quad \forall \l\ne\mu,
\eqa
or equivalently:
\bq
\bra{\psi_{n}}E_\mu\ket{\psi_{n}}\gg\bra{\psi_{n}}E_\l\ket{\psi_{n}},\quad
\quad \forall \l\ne\mu. 
\eq
The probability $p_\mu$ for this particular eigenvalue $e^{i\mu}$ to
come out is determined by the initial state $\ket{\psi_0}$: 
\bq
p_\mu=\bra{\psi_0}E_\mu\ket{\psi_0}.
\eq

We will now assume that there are only two distinct eigenvalues
$e^{i\l_1}$ and $e^{i\l_2}$; later on, we will give a generalization
for cases with more than two eigenvalues. As the following part of the
proof is purely mathematical, it is convenient to use an abstract
notation which does not directly relate to the thought-experimental
setup of the interference experiments. Let us introduce the functions
$A(i)$, $B(i)$, $p(i_1,\ldots,i_n)$ and the constants $\alpha^2$,
$\beta^2$, by making the following identifications: 
\bqa
A(i)&\equiv& P_{e^{i\l_1}}[D_i]\\
B(i)&\equiv& P_{e^{i\l_2}}[D_i]\\
p(i_1,\ldots,i_n)&\equiv&p_n(D_{i_1},\ldots,D_{i_n})\\
\alpha^2&\equiv&\bra{\psi_0}E_{\l_1}\ket{\psi_0}\\
\beta^2&\equiv&\bra{\psi_0}E_{\l_2}\ket{\psi_0}.
\eqa
All are non-negative, and can be easily shown to obey the following
relations: 
\[
\sum_i A(i)=1,\quad\quad \sum_i B(i)=1,\quad\quad\alpha^2+\beta^2=1,
\]
\[
p(i_1,\ldots,i_n)=\alpha^2 A(i_1)\ldots A(i_n) + \beta^2 B(i_1)\ldots B(i_n),
\]
\[
\sum_{i_1,\ldots,i_n}p(i_1,\ldots,i_n)=1.
\]

Next, and this is the pivot of the proof, we define a \emph{new}
probability distribution function $P_n(z)$ of a real-valued variable
$z$, which in turn is defined through the relation: 
\bq
e^z=\frac{\bra{\psi_n}E_{\l_1}\ket{\psi_n}}{\bra{\psi_n}
E_{\l_2}\ket{\psi_n}}, 
\eq
i.e., the probability that after the $n$th run the expectation value
of $E_{\l_1}$ is equal to that of $E_{\l_2}$ times the exponent of
$z$. The formal definition of $P_n(z)$ involves a Dirac
delta-function: 
\bq
P_n(z)=\sum_{i_1,\ldots,i_n}p(i_1,\ldots,i_n)\,\delta\!
\left(z-\ln\case{\bra{\psi_n}E_{\l_1}\ket{\psi_n}}
{\bra{\psi_n}E_{\l_2}\ket{\psi_n}}\right). 
\eq
As a probability distribution $P_n(z)$ should be normalized, as it is:
\bq
\int^{\infty}_{-\infty}P_n(z)dz=1.
\eq
Convergence, i.e., locking, is achieved when either
$\bra{\psi_n}E_{\l_1}\ket{\psi_n}\gg\bra{\psi_n}E_{\l_2}\ket{\psi_n}$
or
$\bra{\psi_n}E_{\l_2}\ket{\psi_n}\gg\bra{\psi_n}E_{\l_1}\ket{\psi_n}$,
which means that $P_n(z)$ needs to become zero near $z=0$, because the
area under $P_n(z)$ around $z=0$ stands for the probability that
$\bra{\psi_n}E_{\l_1}\ket{\psi_n}$ is approximately equal to
$\bra{\psi_n}E_{\l_2}\ket{\psi_n}$. 

If we use that:
\[
\ln \frac{\bra{\psi_n}E_{\l_1}\ket{\psi_n}}
{\bra{\psi_n}E_{\l_2}\ket{\psi_n}}= 
\ln\frac{\alpha^2}{\beta^2}+\ln\frac{A(i_1) \ldots
A(i_n)}{B(i_1)\ldots B(i_n)}, 
\]
then $P_n(z)$ can be conveniently written as the sum of two functions
that are scaled by $\alpha^2$ and $\beta^2$: 
\bq\label{eq:Pn_sum}
P_n(z)=\alpha^2 P^A_n(z-\ln\case{\alpha^2} {\beta^2})+\beta^2
P^B_n(z-\ln\case{\alpha^2}{\beta^2}). 
\eq
These two functions $P^A_n(z)$ and $P^B_n(z)$ are independent of
$\alpha^2$ and $\beta^2$: 
\bq\label{eq:app:PAn}
P^A_n(z)=\sum_{i_1,\ldots,i_n} A(i_1)\ldots
A(i_n)\,\delta\!\left(z-\ln\case{A(i_1)\ldots A(i_n)}{B(i_1)\ldots
B(i_n)}\right), 
\eq
\bq\label{eq:app:PBn}
P^B_n(z)=\sum_{i_1,\ldots,i_n} B(i_1)\ldots
B(i_n)\,\delta\!\left(z-\ln\case{A(i_1)\ldots A(i_n)}{B(i_1)\ldots
B(i_n)}\right), 
\eq
and can both be regarded as probability distribution functions, since:
\bq
\int^{\infty}_{-\infty}P^A_n(z)dz=1,\quad\quad
\int^{\infty}_{-\infty}P^B_n(z)dz=1. 
\eq
Next, we will calculate the expectation values of $z$ and $z^2$ with
respect to $P^A_n(z)$ and $P^B_n(z)$. 

The expectation value $\langle z \rangle^A_{1}$ of $z$ for $P^A_1(z)$ is:
\bqa
\langle z \rangle^A_{1}=\int^{\infty}_{-\infty}P^A_1(z)\,z\,
dz&=&\sum_i A(i) \ln \frac{A(i)}{B(i)}\nn\\ 
&\equiv& m_A,
\eqa
where $m_A>0$ because $A(i)\ne B(i)$ and:%
\footnote{$A(i)$ unequal to $B(i)$ means $P_{e^{i\l_1}}[D_i]\ne
P_{e^{i\l_2}}[D_i]$, which is usually true unless $Q$ is zero, in
which case there is no observable interference, and thus the whole
experiment would be uninteresting. 
}%
\[
a\ln\frac{a}{b}+b\ln\frac{b}{a}=(a-b)(\ln a-\ln b)> 0, \quad\quad a\ne b.
\]
The expectation value of $z^2$ is given by:
\bqa
\langle z^2 \rangle^A_{1}=\int^{\infty}_{-\infty}P^A_1(z)\,z^2\,
dz&=&\sum_i A(i) \ln^2 \frac{A(i)}{B(i)}\nn\\ 
&\equiv& s_A^2+m_A^2,
\eqa
which also defines $ s_A^2$. We can define $m_B$ and $ s_B^2$ in a
similar way: 
\bqa
\langle z \rangle^B_{1}=\int^{\infty}_{-\infty}P^B_1(z)\,z\,
dz&=&\sum_i B(i) \ln \frac{A(i)}{B(i)}\nn\\ 
&\equiv&-m_B<0,
\eqa
\bqa
\langle z^2 \rangle^B_{1}=\int^{\infty}_{-\infty}P^A_1(z)\,z^2\,
dz&=&\sum_i B(i) \ln^2 \frac{A(i)}{B(i)}\nn\\ 
&\equiv& s_B^2+m_B^2.
\eqa

If we now calculate the expectation values $\langle z \rangle^A_{n}$,
$\langle z^2 \rangle^B_{n}$, etc., for arbitrary $n$ by plugging in
eqs.~(\ref{eq:app:PAn}) and (\ref{eq:app:PBn}), we find these depend
on $n$ in the following way: 
\bq
\langle z \rangle^A_{n}=\int^{\infty}_{-\infty}P^A_n(z)\,z\,
dz=n\langle z \rangle^A_1=nm_A. 
\eq
\bq
\langle z^2 \rangle^A_{n}=n s_A^2+n^2m^2_A
\eq
\bq
\langle z \rangle^B_{n}=nm_B,\quad\quad\langle z^2 \rangle^B_{n}=n
s_B^2+n^2m^2_B 
\eq
The expectation values of $z$ and $z^2$ tell a great deal over the
general form of $P_n^A(z)$ and $P^B_n(z)$, by considering the mean
$\mu=\langle z\rangle$ and the variance $\sigma^2=\langle
z^2\rangle-\langle z\rangle^2$ of $P_n^A(z)$ and $P^B_n(z)$: 
\bq
\mu_{An}=\langle z \rangle^A_{n}=n\, m_A,
\eq
\bq
 \sigma^2_{An}=\langle z^2 \rangle^A_{n}-\left(\langle z
\rangle^A_{n}\right)^2=n\,s_A^2, 
\eq
\bq
\mu_{Bn}=-n\, m_B,\quad\quad \sigma^2_{Bn}=n\, s_B^2.
\eq
Both the mean $\mu$ and the variance $\sigma^2$ of $P^A_n(z)$ and
$P^B_n(z)$ scale linearly with $n$. The standard-with $\sigma$,
however, is given by the square root of the variance $\sigma^2$, and
thus scales with $\sqrt{n}$. This means that if we compare $P^A_1(z)$
and $P^A_{100}(z)$, the mean of $P^A_{100}(z)$ is a hundred times
greater than that of $P^A_1(z)$, while the width has only increased by
a factor ten. The total area under both $P^A_1(z)$ and $P^A_{100}(z)$
remains equal to one. 

So, $P^A_n(z)$ and $P^B_n(z)$ are both probability distribution
functions, of which the peak `runs away' from the origin when $n$
increases, and these peaks run harder than they broaden. This behavior
is in fact sufficient as proof. For let us look at $P_n(z)$,
eq.~(\ref{eq:Pn_sum}), which is after all the probability distribution
function we are physically interested in, for increasing $n$. The area
under $P_n(z)$ for large values of $z$ indicates the probability that
after $n$ runs $\bra{\psi_{n}}E_{\l_1}\ket{\psi_{n}}=1$; the area
under $P_n(z)$ for large negative $z$ is the probability that
$\bra{\psi_{n}}E_{\l_2}\ket{\psi_{n}}=1$. By now, we know that
$P_n(z)$ is completely dominated by the $P^A_n(z)$-part for large $z$
and by the $P^B_n(z)$-part for large negative $z$, because of the way
the means and widths scale with $n$. 

Nevertheless, let us specify ``large $z$'' by choosing a fixed
$z'>0$,%
\footnote{To be precise, if $z\ge z'$, this means that
$\bra{\psi_{n}}E_{\l_1}\ket{\psi_{n}}\ge\frac{1}{1+e^{-z'}}$. So, if
$z'=25$, then $e^{z'}\approx10^{10}$, and thus
$\bra{\psi_{n}}E_{\l_1}\ket{\psi_{n}}\ge 1 - 10^{-10}$: the state is
pretty much projected, and we can regard $z'=25$ as large. 
}%
  and calculate some $z'$-dependent areas under $P_n(z)$ when $n$ goes
to infinity: 
\bq
\lim_{n\to\infty}\int^{z'}_{-z'} P_n(z)\,dz=0,
\eq
\bq
\lim_{n\to\infty}\int^{\infty}_{z'} P_n(z)\,dz=\alpha^2,
\eq
\bq
\lim_{n\to\infty}\int_{-\infty}^{-z'} P_n(z)\,dz=\beta^2.
\eq
Clearly, the results of these integrals do not depend on our arbitrary
choice of $z'$. And so, the projection is proved, i.e., at the end of
the experiment the system is locked with complete certainty, including
the correct probabilities for each possible outcome, since
$\alpha^2=p_{\l_1}$ and $\beta^2=p_{\l_2}$. 

The proof for more than two eigenvalues is a simple
generalization. For the case of three eigenvalues, $P_n$ depends on
two variables $z$ and $w$, where $P_n(z,w)$ then stands for the
probability that at the same time both: 
\bq
e^z=\frac{\bra{\psi_n}E_{\l_1}\ket{\psi_n}}{\bra{\psi_n}E_{\l_2}
\ket{\psi_n}},\quad\quad 
e^w=\frac{\bra{\psi_n}E_{\l_1}
\ket{\psi_n}}{\bra{\psi_n}E_{\l_3}\ket{\psi_n}}. 
\eq
$P_n(z,w)$ can now be written as a sum of three other scaled
probability distribution functions $P^A_n(z,w)$, $P^B_n(z,w)$, and
(because of the third eigenvalue) $P_n^C(z,w)$, which all `run away'
from the origin with increasing $n$. This procedure can be easily
generalized for an arbitrary, but finite, number of eigenvalues. 

For the many-to-one experiment, the recipe of the proof is simply:
replace the eigenvalues and projector operators by their appropriate
counterparts, i.e., replace each $e^{i\l}$ by $\kappa$, each
$P_{e^{i\l}[D_i]}$ by $P_\kappa[D_i]$, and each $E_\l$ by
$E_\kappa$. Although $\ket{\psi_n}$ in the many-to-one experiment is
very different from $\ket{\psi_n}$ in the one-to-one experiment, i.e.,
an element of a very different vectorspace, and the same applies to
$E_\l$ and $E_\kappa$, we can nevertheless in the many-to-one
experiment come up with an expression similar to (\ref{eq:En_expec1}): 

\bq\label{eq:En_expec2}
\bra{\psi_{n}}E_\kappa\ket{\psi_{n}}=\frac{1}{K}P_{\kappa}[D_{i_1}]\ldots
P_{\kappa}[D_{i_n}]\,\bra{\psi_0}E_\kappa\ket{\psi_0}. 
\eq
We arrive at (\ref{eq:En_expec2}) by determining
$\bra{\psi_{j+1}}E_\kappa\ket{\psi_{j+1}}$ in terms of
$\bra{\psi_{j}}E_\kappa\ket{\psi_{j}}$, by doing some rewriting of
$\bra{\psi_{j+1}}\R^2_{j+2}\ket{\psi_{j+1}}$: 
\bqa
\bra{\psi_{j+1}}\R^2_{j+2}\ket{\psi_{j+1}}&=&\sum_\kappa \kappa\,
\bra{\psi_{j+1}}E_\kappa\ket{\psi_{j+1}}\\ 
&=&\frac{1}{K'}\sum_\kappa \kappa\,
P_\kappa[D_{i_j}]\,\bra{\psi_{j}}E_\kappa\ket{\psi_{j}}. 
\eqa
From eq.~(\ref{eq:En_expec2}) forward, the proof of the one-to-one and
many-to-one experiments is exactly the same. 

The functions $P^A_n(z)$ and $P^B_n(z)$ have other interesting
properties as well, for instance $P^A_n(z)=e^z P^B_n(z)$, but also
$P^A_2=P^A_1\circ P^A_1$ while $P_2\ne P_1\circ P_1$. Furthermore the
sums over $i$, where $i$ could be 1 or 2 corresponding to detector
$D_1$ or $D_2$, can easily be generalized to more detectors, i.e.,
sums where $i$ can take on more values, and can also be replaced by
integrals for setups with a `continuous spectrum of detectors'. These
aspects we will not discuss here, as they lie outside the current
scope; see \cite{BJOscriptie} for some more details.


\begin{thebibliography}{10}

\bibitem{wilczek}
F.~Wilczek,
\newblock Quantum mechanics of fractional spin particles,
\newblock Phys. Rev. Lett. {\bf 49}, 957 (1982).

\bibitem{decoherence}
 {\em Quantum coherence and decoherence}, No.~1969 in {\em Proceedings of
  the Royal Society of London SERIES A}, edited by D.~DiVincenzo (Royal
  Society, London, 1998), pp.\ 257--486.

\bibitem{preskill}
J.~Preskill,
\newblock Fault-tolerant quantum computation,
\newblock in {\em Introduction to quantum computation and information}, edited
  by H.-K.~Lo, S.~Popescu, and T.~Spiller, World Scientific, Singapore, 1998,
  quant-ph/9712048.

\bibitem{ekert1}
A.~Ekert {\em et~al.},
\newblock Geometric Quantum Computing,
\newblock Journal of Modern Optics {\bf 47}, 2501 (2000), quant-ph/0004015.

\bibitem{lloyd}
S.~Lloyd,
\newblock Quantum computation with abelian anyons,
\newblock quant-ph/0004010.

\bibitem{kitaev}
A.~Kitaev,
\newblock Fault-tolerant quantum computation by anyons,
\newblock quant-ph/9707021.

\bibitem{freedman1}
M.~H. Freedman, A.~Kitaev, M.~J. Larsen, and Z.~Wang,
\newblock Topological Quantum Computation,
\newblock quant-ph/0101025.

\bibitem{freedman2}
M.~H. Freedman,
\newblock $P/NP$, and the quantum field computer,
\newblock Proc. Natl. Acad. Sci. USA {\bf 95}, 98 (1998).

\bibitem{bais0}
F.~A. Bais,
\newblock Flux metamorphosis,
\newblock Nucl. Phys. {\bf B170}, 32 (1980).

\bibitem{bais1}
F.~A. Bais, P.~van Driel, and M.~de~Wild~Propitius,
\newblock Quantum symmetries in discrete gauge theories,
\newblock Phys. Lett. {\bf B280}, 63 (1992), hep-th/9203046.

\bibitem{bais2}
M.~de~Wild~Propitius and F.~A. Bais,
\newblock Discrete gauge theories,
\newblock in {\em Particles and fields}, edited by G.~Semenoff and L.~Vinet,
  CRM series in Mathematical Physics, pp. 353--439, New York, 1998,
  Springer-Verlag, hep-th/9511201.

\bibitem{ekert2}
E.~Sjoqvist {\em et~al.},
\newblock Geometric phases for mixed states in interferometry,
\newblock Physical Review Letters {\bf 85}, 2845 (2000), quant-ph/0005072.

\bibitem{verlinde}
E.~Verlinde,
\newblock A note on braid statistics and the nonabelian Aharonov-Bohm effect,
\newblock in {\em Proceedings of the International Colloquium on Modern Quantum
  Field Theory}, edited by S.~Das, pp. 450--461, Singapore, 1991, World
  Scientific.

\bibitem{lo}
H.-K. Lo and J.~Preskill,
\newblock NonAbelian vortices and nonAbelian statistics,
\newblock Phys. Rev. {\bf D48}, 4821 (1993), hep-th/9306006.

\bibitem{fqhe1}
C.~Nayak and F.~Wilczek,
\newblock $2n$ quasiholes realize $2^{n-1}$-dimensional spinor braiding statistics
 in paired quantum Hall states,
\newblock Nucl. Phys. {\bf B479}, 529 (1996), cond-mat/9605145.

\bibitem{fqhe2}
J.~K. Slingerland and F.~A. Bais,
\newblock Quantum groups and nonabelian braiding in quantum Hall systems,
\newblock cond-mat/0104035.

\bibitem{zeilinger:1981}
A.~Zeilinger,
\newblock General Properties of Lossless Beam Splitters in Interferometry,
\newblock Amer. J. Phys. {\bf 49}, 882 (1981).

\bibitem{silverman:mystery}
M.~P. Silverman,
\newblock {\em More than One Mystery: Explorations in Quantum Interference}
  (Springer-Verlag, New York, 1995).

\bibitem{BJOscriptie}
B.~J. Overbosch,
\newblock The Entanglement and Measurement of non-Abelian
 Anyons as an Approach to Quantum Computation,
\newblock Master's thesis, University of Amsterdam, 2000.

\end{thebibliography}

\end{document}